\documentclass[aps,prc,reprint,showpacs,superscriptaddress,nofootinbib]{revtex4-1}

\usepackage[T1]{fontenc}
\usepackage[utf8]{inputenc}
\usepackage{graphicx,color,rotating,pifont}
\usepackage{amsmath,amssymb}
\usepackage{mathtools}
\usepackage{siunitx}
\usepackage{bm}
\usepackage{ae}
\usepackage{dcolumn}
\usepackage{txfonts}
\usepackage{tensor}
\usepackage{braket}
\usepackage{booktabs}
\usepackage{xspace}
\usepackage{soul}
\setstcolor{red}
\setul{}{.2ex} 

\usepackage{tikz}
\usetikzlibrary{positioning,calc,shapes,decorations.markings,decorations.pathmorphing}
\tikzset{asg/.cd,
  omega-vertex/.style={circle,solid,draw=black,fill=white,minimum size=5pt, inner sep=0pt},
  dbd-vertex/.style={coordinate},
  pline/.style={thick, postaction={decorate}, decoration={markings, mark=at position .5 with {\arrow[xshift=2pt]{stealth}}}},
  hline/.style={thick, postaction={decorate}, decoration={markings, mark=at position .5 with {\arrowreversed[xshift=-2pt]{stealth}}}},
  shift arrow/.style={/pgf/decoration/transform={xshift=#1}},
  shift arrow/.default=-2pt,
  dbd-2b/.style={decorate, decoration=snake},
  omega-2b/.style={densely dashed},
  neutron/.style={draw=blue},
  proton/.style={draw=red},
}

\usepackage{hyperref}

\begingroup%
\makeatletter%
\catcode`\^^M=12\relax%
\def\activereturnfork#1{%
  \endgroup%
  \newcommand\activereturnfork[1]{%
    \forkactivereturn##1{}#1{##1}^^M^^M%
  }%
  \@ifdefinable\forkactivereturn{%
    \long\def\forkactivereturn##1#1##2##3^^M^^M{##2}%
  }%
}%
\catcode`\^^M=13\relax%
\activereturnfork{^^M}%
\makeatother

\DeclareMathSymbol{\NS}{\mathord}{AMSb}{"4E}

\DeclareSIUnit{\fm}{\femto\meter}
\DeclareSIUnit{\MeVc}{\MeV\per\text{\ensuremath{c}}}

\newcommand{\matrixe}[3]{\ensuremath{\braket{{#1}|{#2}|{#3}}}}

\newcommand{\op}[1]{\ensuremath{#1}}
\newcommand{\adj}[1]{\ensuremath{{{#1}}^{\dag}}}

\newcommand{\dd}{\ensuremath{\mathrm{d}}}

\renewcommand{\vec}[1]{\ensuremath{\bm{#1}}}

\newcommand{\aO}{\ensuremath{\op{a}}}

\newcommand{\aaO}{\ensuremath{\adj{\op{a}}}}

\newcommand{\AO}{\ensuremath{\op{A}}}

\newcommand{\HO}{\ensuremath{\op{H}}}

\newcommand{\UUO}{\ensuremath{\adj{\op{U}}}}

\newcommand{\Gj}[6]{ \begin{Bmatrix}
   #1 & #2 & #3 \\
   #4 & #5 & #6 
  \end{Bmatrix}}

\newcommand{\totd}[2]{\ensuremath{ \frac{\dd {#1}} {\dd {#2}} }}

\newcommand{\eMax}{\ensuremath{e_{\text{max}}}}

\newcommand{\NMax}{\ensuremath{N_{\text{max}}}}

\newcommand{\beq}{\begin{equation}}
\newcommand{\eeq}{\end{equation}}
\newcommand{\beqn}{\begin{eqnarray}}
\newcommand{\eeqn}{\end{eqnarray}}
\newcommand{\bsub}{\begin{subequations}}
\newcommand{\esub}{\end{subequations}}
\newcommand{\bpm}{\begin{pmatrix}}
\newcommand{\epm}{\end{pmatrix}}

\newcommand{\nord}[1]{\ensuremath{\mathinner{\mathopen{:}{#1}\mathclose{:}}}}

\definecolor{FGViolet}{rgb}{0.61,0.32,0.61}
\definecolor{FGDarkBlue}{rgb}{0,0,0.6}
\definecolor{FGBlue}{rgb}{0,0,0.8}
\definecolor{FGLightBlue}{rgb}{0.2, 0.6, 0.8}
\definecolor{FGGreen}{rgb}{0.2,0.7,0.2}
\definecolor{FGLightGreen}{rgb}{0.4,1,0.4}
\definecolor{FGYellow}{rgb}{1,0.95,0}
\definecolor{FGOrange}{rgb}{0.95,0.5,0.1}
\definecolor{FGRed}{rgb}{0.8,0,0}
\definecolor{FGWhite}{rgb}{1,1,1}
\definecolor{FGLightGray}{rgb}{0.8,0.8,0.8}
\definecolor{FGGray}{rgb}{0.5,0.5,0.5}
\definecolor{FGDarkGray}{rgb}{0.3,0.3,0.3}
\definecolor{FGBlack}{rgb}{0,0,0}

\begin{document}

\title{{\em Ab initio} benchmarks of neutrinoless double beta decay in light nuclei with a chiral Hamiltonian}

\author{J. M. Yao}  
\email{yaoj@frib.msu.edu}
 \affiliation{Facility for Rare Isotope Beams, Michigan State University, East Lansing, Michigan 48824-1321, USA}
  
\author{A. Belley}  
\email{abelley@triumf.ca}
\affiliation{TRIUMF 4004 Wesbrook Mall, Vancouver BC V6T 2A3, Canada}%
\affiliation{Department of Physics, McGill University, 3600 Rue University, Montr\'eal, QC H3A 2T8, Canada}%
\affiliation{Department of Physics \& Astronomy, University of British Columbia, Vancouver, British Columbia V6T 1Z1, Canada}

 \author{R. Wirth}  
 \email{wirth@frib.msu.edu}
  \affiliation{Facility for Rare Isotope Beams, Michigan State University, East Lansing, Michigan 48824-1321, USA}

\author{T. Miyagi}%
\affiliation{TRIUMF 4004 Wesbrook Mall, Vancouver BC V6T 2A3, Canada}%

\author{C. G. Payne}%
\altaffiliation[Present address: ]{Institut f\"ur Kernphysik and PRISMA$^+$ Cluster of Excellence, Johannes Gutenberg-Universit\"at at Mainz, 55128 Mainz, Germany}
\affiliation{TRIUMF 4004 Wesbrook Mall, Vancouver BC V6T 2A3, Canada}%
\affiliation{Department of Physics \& Astronomy, University of British Columbia, Vancouver, British Columbia V6T 1Z1, Canada}

\author{S. R. Stroberg}%
\affiliation{Department of Physics, University of Washington, Seattle, WA 98195, USA}

 \author{H. Hergert} 
  \affiliation{Facility for Rare Isotope Beams, Michigan State University, East Lansing, Michigan 48824-1321, USA}
 \affiliation{Department of Physics \& Astronomy, Michigan State University, East Lansing, Michigan 48824-1321, USA}
  
\author{J. D. Holt}%
\affiliation{TRIUMF 4004 Wesbrook Mall, Vancouver BC V6T 2A3, Canada}%
\affiliation{Department of Physics, McGill University, 3600 Rue University, Montr\'eal, QC H3A 2T8, Canada}%

\date{\today}

\begin{abstract}
 We report {\em ab initio} benchmark calculations of nuclear matrix elements (NMEs) for  neutrinoless double-beta  ($0\nu\beta\beta$) decays in light nuclei with mass number ranging from $A=6$ to $A=22$.
 We use the transition operator derived from light-Majorana neutrino exchange and evaluate the NME with three different methods: two variants of  in-medium similarity renormalization group (IMSRG) and importance-truncated no-core shell model (IT-NCSM).
 The same  two-plus-three-nucleon interaction from chiral effective field theory is employed, and both isospin-conserving ($\Delta T=0$) and isospin-changing ($\Delta T=2$) transitions are studied. We compare our resulting ground-state energies and NMEs to those of recent {\em ab initio} no-core shell model and coupled-cluster calculations, also with the same inputs.
 We show that the NMEs of $\Delta T=0$ transitions are in good agreement among all calculations, at the level of \SI{10}{\percent}. For $\Delta T=2$, relative deviations are more significant in some nuclei. The comparison with the exact IT-NCSM result allows us to analyze these cases in detail, and indicates the next steps towards improving the IMSRG-based approaches. 
 The present study clearly demonstrates the power of consistent cross-checks that are made possible by {\em ab initio} methodology. This capability is crucial for providing meaningful many-body uncertainties in the NMEs for the $0\nu\beta\beta$ decays in heavier candidate nuclei, where quasi-exact benchmarks are not available.

\end{abstract}
 
\maketitle
 
\section{Introduction}
   
 Double-beta decay is a second-order weak transition which manifests  itself in the ``low-energy" environment of atomic nuclei as two neutrons in a parent nucleus ($A, Z$) decaying into two protons in a daughter nucleus ($A, Z+2$) via the emission of two electrons, with or without accompanying two-neutrino emission. This corresponds to the two decay modes: two-neutrino double-beta ($2\nu\beta\beta$) decay and neutrinoless double-beta ($0\nu\beta\beta$) decay, respectively. The $2\nu\beta\beta$ decay mode~\cite{Goeppert-Mayer:1935} is allowed in the Standard Model of particle physics and has been observed in several atomic nuclei with half-lives ranging from $10^{19}-10^{21}$ years~\cite{Saakyan:2013}. In contrast, $0\nu\beta\beta$ decay is a hypothetical lepton-number-violating process forbidden in the Standard Model of particle physics~\cite{Furry:1939}. The hunt for  $0\nu\beta\beta$ decay is of particular importance, as its observation would demonstrate the Majorana nature of neutrinos and provide a key ingredient for generating the matter-antimatter asymmetry in the Universe. Currently, the best half-life lower limit (>$10^{25}$ years) is achieved in the experiments on $^{136}$Xe~\cite{KamLAND-Zen2016,EXO-200:2019}, $^{76}$Ge~\cite{GERDA2018} and $^{130}$Te~\cite{CUORE2020}. The null $0\nu\beta\beta$ decay signal from current experiments provides a constraint on the upper limits of effective neutrino mass $\langle m_{\beta\beta}\rangle$ if the decay is mediated by the exchange of light-Majorana neutrinos. In this scenario, the next-generation tonne-scale experiments with half-life sensitivity up to $10^{28}$ years after a few years of running are expected to provide a definite answer on the mass hierarchy of neutrinos based on our current knowledge on the nuclear matrix element (NME) $M^{0\nu}$ of $0\nu\beta\beta$ decay.

Accurate theoretical values of the NMEs are vital for the design and interpretation of future experiments. Based on the standard light-Majorana neutrino-exchange mechanism, various nuclear models have been applied to compute the NMEs of candidate $0\nu\beta\beta$ decays~\cite{Menendez:2009,Rodriguez:2010,Barea:2013,Mustonen:2013,Holt:2013,Kwiatkowski:2014,Song:2014,Yao:2015,Hyvarinen:2015,Horoi:2016,Song:2017,Jiao:2017,Yoshinaga:2018,Fang:2018,Rath:2019,Terasaki:2019,Coraggio:2020,Deppisch:2020ztt}. The discrepancy among predictions is up to a factor of about three, causing an uncertainty at the level of an order of magnitude in the half-life for a given value of the neutrino mass. Resolving this discrepancy has been one of the most significant objectives in the nuclear theory community~\cite{Menendez:2014,Menendez:2016,Engel:2017,LongRangePlan2015}.
 
 A first-principles calculation of the NMEs of candidate $0\nu\beta\beta$ decays is crucial as the theoretical uncertainty from both many-body wavefunctions and transition operators must be under control. This calculation is tremendously challenging, if not impossible, for \emph{quasi-exact} many-body approaches. It is, however, within the reach of some \emph{ab initio} methods with systematically improvable approximations, such as coupled-cluster (CC) theory~\cite{Hagen:2014} and the in-medium similarity renormalization group (IMSRG)~\cite{Hergert:2016jk}, where the computational complexity scales polynomially with nuclear size. Within this framework, three \emph{ab initio} methods, i.e., the in-medium generator coordinate method (IM-GCM)~\cite{Yao:2020}, valence-space IMSRG (VS-IMSRG)~\cite{Belley2020},
 and CC theory with singles and doubles plus leading-order triples excitations (CCSDT1)~\cite{Novario2020},
 have recently been used to calculate the $0\nu\beta\beta$-decay NME of $^{48}$Ca, or those of even heavier candidates $^{76}$Ge and $^{82}$Se \cite{Belley2020}, starting from a two-nucleon-plus-three-nucleon (NN+3N) interaction constructed from chiral effective field theory (EFT), where the same transition operator derived from the light-Majorana neutrino exchange was adopted. For a recent summary of these results, see Ref.~\cite{Yao:2020Science}. Since there is no \emph{exact} solution or experimental data available that can be used to validate the NMEs of candidate decays, synthetic benchmarks, i.e., $0\nu\beta\beta$ decays that are energetically forbidden or occur in competition with single $\beta$ decay, may provide a unique way of cross-checking among different \emph{ab initio} methods. Fortunately, light nuclei are within the reach of several quasi-exact approaches, the results of which are valuable to validate the adopted many-body approximations and the usefulness of the cross-checking strategy among  different \emph{ab initio} methods.

 Following this philosophy, significant progress has been made in the calculation of the NMEs of $0\nu\beta\beta$ decays in a set of light nuclei using nuclear wavefunctions from several many-body approaches. In Refs.~\cite{Pastore:2018,Wang:2019}, the NMEs for $^{6,8,10}$He and $^{10,12}$Be were calculated with a quantum Monte Carlo (QMC) method based on the NN interaction AV18 \cite{AV18:1995} and the 3N interaction IL7 \cite{IL7:2008}. The same QMC calculation was later carried out based  on the local Norfolk chiral NN+3N potentials \cite{Baroni:2018} for the NMEs of $\nuclide[6]{He}$ and $\nuclide[12]{Be}$ \cite{Cirigliano:2019PRC}.  In Ref.~\cite{Basili2020} the NME for $^{6}$He was calculated with a no-core shell model (NCSM)~\cite{Barrett:2013} based on a similarity renormalization group (SRG)-softened chiral NN interaction~\cite{Basili2020}. In Ref.~\cite{Novario2020}, the NMEs of $^{6,8,10}$He, $^{14}$C and $^{22}$O were also calculated with the NCSM based on a chiral NN+3N interaction.

 In this paper, we present \emph{ab initio} calculations of the NMEs of $0\nu\beta\beta$ decay (including both $\Delta T=0$ and $\Delta T=2$ transitions) in several light nuclei with mass number ranging from $A=6$ to $A=22$ using two variants of the IMSRG---the VS-IMSRG \cite{Stroberg:2016,Stroberg:2017,Stroberg:2019} and the IM-GCM \cite{Yao:2018wq,Yao:2020}---and the importance-truncated (IT) NCSM \cite{Roth:2009} starting from the same chiral NN+3N interaction, aiming to benchmark the possible errors in the ground-state energies and the NMEs of $0\nu\beta\beta$ decay.   
 
We organize the paper as follows: In Sec.~\ref{sec:approaches}, we review the framework of IT-NCSM very briefly.
Then, we introduce the formalism of the two IMSRG variants in detail, which provide inputs of effective interactions and $0\nu\beta\beta$ decay operators for subsequent conventional nuclear many-body calculations.
In Sec.~\ref{results}, both the ground-state energies and NMEs from the three calculations are compared to reported calculations~\cite{Novario2020} using the same operators, and we analyze the renormalization effect on the distribution of the NMEs in coordinate space.
A summary and perspective is given in Sec.~\ref{summary}.

 \section{\label{sec:approaches}\emph{Ab initio} approaches}
 
 \subsection{Hamiltonian}
 We employ an intrinsic nuclear $A$-body Hamiltonian containing NN+3N interactions from chiral EFT, 
\begin{align}
\label{Eq:H}
H_0 &= \sum_{i<j} \dfrac{(\vec{p}_i-\vec{p}_j)^2}{2mA} + \sum_{i<j} V_{ij}^{(2N)} +  \sum_{i<j<k} V_{ijk}^{(3N)}\nonumber\\
 &=\left(1-\dfrac{1}{A}\right)\sum_i \dfrac{\vec{p}^2_i}{2m}-\dfrac{1}{A}\sum_{i<j}\dfrac{(\vec{p}_i\cdot\vec{p}_j)}{m} \nonumber\\
&\hphantom{{}={}}+\sum_{i<j}V^{(2N)}_{ij}+\sum_{i<j<k}V^{(3N)}_{ijk} 
\end{align}
which in second-quantization form reads,
\begin{align} 
 H_0 &= \sum_{pq} t^p_{q} A^p_q  + \dfrac{1}{4}\sum_{pqrs} V^{pq}_{rs}
A^{pq}_{rs}
+\dfrac{1}{36}\sum_{pqrstu}W^{pqr}_{stu} A^{pqr}_{stu}, 
\end{align}
where $t$, $V$, and $W$ represent one-body, two-body and three-body interaction matrices, respectively. We introduce strings of creation and annihilation operators as
\beq
  A^{pqr\ldots}_{stu\ldots} = a^\dagger_pa^\dagger_qa^\dagger_r\ldots a_u a_t a_s\,,
\eeq 
where the $p, q, \cdots$ index the states of a spherical harmonic-oscillator (HO) basis. The fit of parameters for the NN interaction $V_{ij}^{(2N)}$, carried out at next-to-next-to-next-to leading
order (N$^3$LO) with a momentum cutoff of \SI{500}{\MeVc}, is from~\citet{Entem:2003}. We use the free-space SRG~\cite{Bogner:2010} to evolve the interaction to a resolution scale of $\lambda=\SI{1.8}{\per\fm}$. Following Refs.~\cite{Hebeler:2011,Nogga:2004il}, we construct the 3N interaction $V_{ijk}^{(3N)}$ directly, with a chiral cutoff of $\Lambda=\SI{2.0}{\per\fm}$. We refer to the
Hamiltonian that results as EM$\lambda$/$\Lambda$, i.e., EM1.8/2.0. See Refs.~\cite{Hebeler:2011,Nogga:2004il} for  details.
For the 3N interaction, we discard all matrix elements involving states with $e_1+e_2+e_3>14$, where $e_i=2n_i+\ell_i$ is the number of oscillator quanta in state $i$. The resulting NN+3N Hamiltonian has been shown to accurately reproduce energies up to approximately the tin region, while systematically underpredicting charge radii~\cite{Simonis:2017,Morris:2018,Holt:2019gm}.
 
\subsection{The importance-truncated no-core shell model}
One of the methods we use to calculate the energy and ground-state wavefunctions of the decay partners under consideration is the importance-truncated no-core shell model (IT-NCSM) \cite{Barrett:2013,Roth:2009,Roth:2011}.
We construct a basis of $A$-body Slater determinants $\ket{\phi_i}$ consisting of single-particle HO states with oscillator frequency $\Omega$. To get a finite basis, we limit the HO excitation energy of the basis determinants to $\NMax\hbar\Omega$ relative to the lowest configuration allowed by Pauli's principle.
In this basis the Hamiltonian turns into a sparse matrix $\mathsf{H}$ with matrix elements $\mathsf{H}_{ij} = \braket{\phi_i|H_0|\phi_j}$, and we solve the matrix eigenvalue problem
\beq
  \mathsf{H}\vec{c} = E \vec{c}
\eeq
to get an approximation to the energy and wavefunction of the ground state in terms of the basis of Slater determinants,
\beq
  \ket{\Psi} = \sum_i c_i \ket{\phi_i}. 
\eeq
This approximation improves with increasing $\NMax$, and the energy and wavefunction converge eventually.

The main challenge in NCSM calculations is the rapid growth of the model-space dimension with both $\NMax$ and $A$. This generally limits the range of applicability of the NCSM to $p$-shell nuclei. For heavier nuclei, the dimensions of the model spaces become too large for current supercomputers to handle before the energy converges.

To mitigate the growth, we employ an importance-trun\-cation scheme~\cite{Roth:2009}: Many of the basis states have only a very small coefficient $c_i$ in the expansion of the ground-state wavefunction.
Consequently, removing these states from the model space only has a small effect on the ground-state wavefunction and the energy.
We estimate the coefficient $c_i$ without actually solving the full eigenvalue problem via perturbation theory:
We compute the first-order correction to a reference wavefunction $\ket{\Psi_{\text{ref}}}$,
\beq
  \kappa_i = -\frac{\braket{\phi_i| H |\Psi_{\text{ref}}}}{\Delta E_i} = -\sum_{j \in \mathcal{M}_{\text{ref}}} c_{\text{ref},j} \frac{\braket{\phi_i| H |\phi_j}}{\Delta E_i}.
\eeq
The reference wavefunction is the ground-state wavefunction obtained from an (IT-) NCSM calculation in a reference space $\mathcal{M}_{\text{ref}}$ with smaller $\NMax$.
The energy difference $\Delta E_i$ is taken as the HO excitation energy of the configuration $\ket{\phi_i}$.
The reduction of the model space is achieved by including only states with $\lvert\kappa_i\rvert \geq \kappa_{\text{min}}$.

We recover the expectation values of observables in the full model space by extrapolating the importance threshold $\kappa_{\text{min}}$ to zero.
To that end, we perform the diagonalization for multiple values of $\kappa_{\text{min}}$ and fit a low-order polynomial to the sequence.
The extrapolated value is obtained by evaluating this polynomial at $\kappa_{\text{min}}=0$.
We estimate the uncertainty of the extrapolation by fitting polynomials of higher and lower orders, and by excluding the smallest $\kappa_{\text{min}}$ values from the fit.

\subsection{The in-medium similarity renormalization group}
\label{models} 
\subsubsection{Reference state}
The basic idea of the IMSRG is to introduce a flow equation to gradually decouple a preselected reference state $\ket{\Phi}$ from all other states \cite{Tsukiyama11,Hergert:2016jk,Hergert:2016} or to decouple the offdiagonal elements of the Hamiltonian that are connecting the $P$ valence space and the excluded $Q$ spaces \cite{Tsukiyama:2012,Bogner:2014tg,Stroberg:2016,Stroberg:2017,Stroberg:2019}. In the former case, the reference state becomes the ground state of the evolved Hamiltonian, while in the latter case, an effective Hamiltonian in a specific valence space is obtained. For a given reference state $\ket{\Phi}$, we first normal-order the Hamiltonian (\ref{Eq:H}) using the generalized normal ordering of Kutzelnigg and Mukherjee \cite{Kutzelnigg:1997JCP,Mukherjee:1997CPL,Kong:2010JCP},
\begin{align}
  H_0 &= E + \sum_{pq} f^{p}_{q} \nord{A^p_q}
    + \dfrac{1}{4}\sum_{pqrs} \Gamma^{pq}_{rs} \nord{A^{pq}_{rs}}
    \notag \\
  &\hphantom{{}={}} + \dfrac{1}{36}\sum_{pqrstu} W^{pqr}_{stu}  \nord{A^{pqr}_{stu}}\,.
\label{normal-ordered-H}
\end{align}
  
By definition, the expectation values of normal-ordered operators, indicated by $\nord{A^{p\ldots}_{q\ldots}}$, with respect to the reference state are zero. Thus, the normal-ordered zero-body term corresponds to the reference-state energy $E$, which is given by
\begin{align}
 E=\braket{ \Phi|H|\Phi}
&=\sum_{pq} t^p_{q}\rho^p_q
+\dfrac{1}{4}\sum_{pqrs}  V^{pq}_{rs} \rho^{pq}_{rs}
\nonumber\\&\hphantom{{}={}}
+\dfrac{1}{36}\sum_{pqrstu}W^{pqr}_{stu}\rho^{pqr}_{stu} \,.
\label{H:0b}
\end{align}
The matrix elements of normal-ordered one-body and two-body terms are
\begin{align}
\label{H:1b}
 f^{p}_{q} &= t^{p}_{q}  +  \sum_{rs} V^{pr}_{qs} \rho^r_s
+\dfrac{1}{4}\sum_{rstu} W^{prs}_{qtu}\rho^{rs}_{tu} \,,\\
\label{H:2b}
 \Gamma^{pq}_{rs} &= V^{pq}_{rs} + \sum_{tu} W^{pqt}_{rsu}\rho^{t}_{u}\,.
\end{align}
In Eqs.~\eqref{H:0b}--\eqref{H:2b}, we have introduced the usual density matrices 
\bsub
\begin{align}
\rho^p_q &= \braket{ \Phi | A^{p}_{q} | \Phi}\,,\\
\rho^{pq}_{rs} &= \braket{ \Phi | A^{pq}_{rs} | \Phi}\,,\\
\rho^{pqr}_{stu} &= \braket{ \Phi | A^{pqr}_{stu} | \Phi}\,.
\end{align}
\esub 
Correlations within the reference state are encoded in the corresponding
\emph{irreducible} density matrices (also referred to as cumulants): 
\bsub
\label{cumulants}
\begin{align}
\lambda^p_q &=  \rho^p_q\,, \\
\lambda^{pq}_{rs} &= \rho^{pq}_{rs}  - \mathcal{A}(\lambda^p_r\lambda^q_s)
 = \rho^{pq}_{rs}  - \lambda^p_r\lambda^q_s +  \lambda^p_s\lambda^q_r\,,\\
\lambda^{pqr}_{stu} &= \rho^{pqr}_{stu} - \mathcal{A}(\lambda^p_s\lambda^{qr}_{tu}+\lambda^p_s\lambda^{q}_{t}\lambda^{r}_{u}) \,,
\end{align}
\esub
where the antisymmetrization operator $\mathcal{A}$ generates all distinct permutations of upper indices and lower indices. For an uncorrelated reference state, the two- and higher-body irreducible densities vanish and we recover the usual factorization of many-body density matrices into
antisymmetrized products of one-body density matrices.

 \subsubsection{Flow equations}
 
 The IMSRG decoupling procedure is realized by introducing a set of unitary transformations $U(s)$ onto the Hamiltonian, 
\begin{align}
\label{Hs}
 H(s)
 &= U(s)  H_0 U^\dagger(s), &   U(0)&=1,
\end{align}
where $s$ is the so-called flow parameter.
The operator $U(s)$ represents a continuous set of unitary transformations that drive $H_0$ to a specific form, e.g., by eliminating certain matrix elements or minimizing its expectation value with respect to unitary transformations \cite{Hergert:2016}. Taking the derivative $\dd/\dd s$ of both sides of Eq.~\eqref{Hs} yields the flow equation 
\beqn
\label{flow-H}
\totd{H(s)}{s} = [ \eta(s),  H(s)] \,,
\eeqn 
where we have introduced the anti-Hermitian generator of the transformation,
\beq
 \eta(s)\equiv\totd{U(s)}{s} U^\dagger (s) \,.
\eeq

The flow equation \eqref{Hs} turns into a set of coupled ordinary differential equations (ODEs), derived from Eq.~\eqref{flow-H}, for $f,\Gamma,\dotsc$ \cite{Tsukiyama11,Hergert:2016jk}. Instead, however, one can solve a similar flow equation for the unitary transformation operator $U(s)$,
\beq\label{eq:flow_U}
\totd{U(s)}{s} =   \eta(s)  U(s)\,.
\eeq
Formally, the solution can be written in terms of the $\mathcal{S}$-ordered
exponential
\beq
 U(s) = \mathcal{S}\exp \int^s_0 \dd s'   \eta(s') \,,
\eeq
which is short-hand for the Dyson series expansion of $U(s)$. As demonstrated first in Ref.~\cite{Morris15}, the flow equation for the unitary operator can be reformulated using the Magnus expansion, which stipulates that the Dyson series can be resummed into a proper exponential,
\beq
  U(s)\equiv e^{ \Omega(s)}\,.
\eeq
The flow equation for $U(s)$ then turns into one for the anti-Hermitian operator $\Omega(s)$,
\beq
\label{flow_omega}
\totd{\Omega(s)}{s}=\sum^\infty_{n=0} \dfrac{B_n}{n!} [\Omega(s), \eta(s)]^{(n)} \,, 
\eeq
where we define nested commutators as
\bsub\begin{align}
  \left[ \Omega(s),  \eta(s)\right]^{(0)} &=  \eta(s)\,,\\ 
  \left[  \Omega(s), \eta(s)\right]^{(n)} &= \left[ \Omega(s), \left[  \Omega(s),  \eta(s)\right]^{(n-1)}\right]\,,
\end{align}\esub
and the $\{B_n\}_{n\geq0} = \{ 1,\linebreak[1] -1/2,\linebreak[1] 1/6,\linebreak[1] 0, \dotsc\}$ are the Bernoulli numbers. As discussed in Ref.~\cite{Morris15}, the reformulation of the IMSRG via the Magnus expansion has two major advantages. First, the anti-Hermiticity of $\Omega(s)$ guarantees that $U(s)$ is unitary throughout the flow, even
when low-order numerical ODE solvers are used to integrate
Eq.~\eqref{flow_omega}. Second, it greatly facilitates the evaluation of observables. In the traditional approach, we would need to solve flow equations
for each additional operator \emph{simultaneously} with Eq.~\eqref{flow-H} because of the dynamical nature of the generator, while $\Omega(s)$ allows us to construct arbitrary evolved operators by using the Baker--Campbell--Hausdorff (BCH) formula:
\beq
\label{BCH}
 O(s) 
= e^{\Omega(s)}  O e^{-\Omega(s)} = 
  \sum^\infty_{n=0} \dfrac{1}{n!} [\Omega(s),O ]^{(n)}
\,.
\eeq
In what follows, we introduce the notation $\overline{O}$ for the evolved operator  $O(s)$ as $s\to\infty$.

 \subsection{Marrying the IMSRG with conventional many-body approaches}
 In order to apply the IMSRG to $0\nu\beta\beta$-decay NME, the calculation of which needs the ground-state wavefunctions of two nuclei, we marry it with conventional many-body approaches, i.e., the generator coordinate method (GCM) and valence-space shell model. To avoid dealing with the difficulty of having two different unitary transformations separately for  initial and final nuclei  as discussed in Ref.~\cite{Yao:2018wq}, only one single unitary transformation is introduced in the IMSRG starting from a common reference state for both nuclei, which provides effective operators as inputs for subsequent conventional many-body approaches. In this work, all the operators are truncated up to normal-ordered two-body (NO2B) terms. The IMSRG under the NO2B approximation is referred to as IMSRG(2) subsequently.

 \subsubsection{IM-GCM}
   In the IM-GCM, the reference state $\vert\Phi\rangle$ of the IMSRG(2) is an ensemble of states for both the initial and final nuclei in the decay, which are obtained by projecting the lowest-energy quasiparticle vacua for each nucleus onto good angular momentum and particle number if not mentioned explicitly. The IMSRG(2) starting from such a general multi-reference state or ensemble is referred to as MR-IMSRG(2) \cite{Hergert:2014}, in which the two- and three-body irreducible densities in \eqref{cumulants} are retained.  We choose the Brillouin generator \cite{Hergert:2016} for $\eta(s)$, which is essentially the gradient of the energy. The MR-IMSRG(2) evolution yields a unitary transformation that transforms all the operators, which are then used as inputs of a subsequent GCM calculation. 
   
   The GCM is a general matrix-diagonalization method for carrying out configuration-mixing calculations in the sense that the nuclear many-body wavefunction is expanded in terms of a set of nonorthogonal basis functions, which are generated by the coordinates $Q_i$ \cite{Ring:1980} associated with the multipole moments that characterize nuclear shapes \cite{Rodriguez:2010,Song:2014,Yao:2015}, pairing amplitudes \cite{Vaquero:2013,Hinohara:2014,Jiao:2017}, or rotational frequency \cite{Borrajo:2015} depending on whether or not they are relevant for the physics of interest. The choice of $Q_i$ defines a model space of many-body configurations whose dimension is usually much smaller than NCSM spaces, since many types of (collective) correlations are already built into the basis functions.  Here,  the basis functions  are chosen as axially deformed quasiparticle vacua with projection onto  particle numbers ($N, Z$) and angular momentum $J$, 
   \beq
   \ket{NZ J, \beta_2} = P^{N,Z} P^{J} \ket{\Phi(\beta_2)}.
   \eeq
   The  ground-state wavefunctions of the initial and final nuclei with spin-parity $J^\pi=0^+$ are constructed as
   \beq
   \ket{ \Psi(0^+_1)} = \sum_{\beta_2} f^{J=0}(\beta_2) \ket{ NZ J=0, \beta_2}
   \eeq 
   The states $\ket{\Phi(\beta_2)}$ labeled by axial deformation parameter ($Q_i=\beta_2$) are a set of quasiparticle vacua determined from variation after particle-number projected Hartree--Fock--Bogoliubov (HFB) calculations with a constraint on the nuclear mass quadrupole deformation
   \beq
   Q_{20}=\braket{\Phi(\beta_2) | r^2 Y_{20} | \Phi(\beta_2)},
   \eeq
   where $\beta_2=4\pi Q_{20}/(3AR^2)$, $R=R_0 A^{1/3}$, and $R_0=\SI{1.2}{\fm}$.
   
   The mixing weight $f^{J}(\beta_2)$ is determined from the variational principle, which leads to the Hill--Wheeler--Griffin (HWG) equation \cite{Ring:1980}
   \beq
   \sum_{\beta'_2} \left[ \mathcal{H}^J(\beta_2,\beta'_2) - E^J \mathcal{N}^J(\beta_2,\beta'_2)\right] f^J(\beta'_2) = 0,
   \eeq
   where the norm ($O=1$) and Hamiltonian ($O=H$) kernels $\mathcal{N}$ and $\mathcal{H}$ are defined as
   \beq
   \mathcal{O}^J(\beta_2,\beta'_2)
   =\braket{ NZ J, \beta_2 | O | NZ J, \beta'_2}.
   \eeq
   The solution of the HWG equation provides both the energies and wavefunctions for the ground states of nuclei involved in the $0\nu\beta\beta$-decay calculation.

    It is worth pointing out that the IMSRG(2) and GCM are complementary to each other in the description of nuclear correlations. The former turns out to be powerful to capture dynamic correlations, but ill-suited for collective correlations \cite{Parzuchowski:2017ta}. In contrast, the latter is successful for the studies of nuclear large-amplitude collective motions \cite{Sheikh:2019}, but inappropriate for low-lying states of spherical nuclei if the configurations of noncollective excitations are not included explicitly.  We refer to the combination of the MR-IMSRG(2) and GCM as the IM-GCM.  This approach has already been applied to calculate the NME for $^{48}$Ca based on a phenomenological shell-model Hamiltonian \cite{Yao:2018wq} and a chiral NN+3N interaction \cite{Yao:2020}.
    
 \subsubsection{VS-IMSRG} 
 In the VS-IMSRG(2), we decouple the interaction within a valence space from the remaining configuration space, and then diagonalize the transformed Hamiltonian exactly in the former. To this end, we split the single-particle model space into core ($c$), valence  particle ($v$) and nonvalence particle ($q$) orbitals. The actual shell model calculation for a nucleus with $A$ nucleons is an exact diagonalization of the Hamiltonian matrix in a  subspace of the Hilbert space that is spanned by configurations  of the form
\begin{equation}\label{eq:def_configurations}
  \ket{v_1\ldots v_{A_v}} \equiv \aaO_{v_1}\ldots\aaO_{v_{A_v}}\ket{\Phi_{A_c}}\,,
\end{equation}
where $\ket{\Phi_{A_c}}$ is the wavefunction for a suitable core with $A_c$
nucleons, and the $A_v$ valence nucleons are distributed over the valence
orbitals $v_i$ in all allowed ways.
The matrix representation of the Hamiltonian in the space spanned by these
configurations is
\begin{equation}
  \matrixe{v'_{1}\ldots v'_{A_v}}{\HO}{v_{1}\ldots v_{A_v}}
  = \matrixe{\Phi_{A_c}}{\aO_{v'_{A_v}}\ldots\aO_{v'_1}\HO\aaO_{v_1}\ldots\aaO_{v_{A_v}}}{\Phi_{A_c}}\,.
\end{equation}

Previous studies \cite{Stroberg:2017,Stroberg:2019} have shown that using the core wavefunction $\ket{\Phi_{A_c}}$ as reference neglects important contributions from the valence-space 3N interaction.
Using an ensemble of Slater determinants with the mass number expectation value set to the mass number $A$ of the target nucleus (instead of the single core determinant $\ket{\Phi_{A_c}}$) as reference has proven advantageous.
Hence, we obtain the reference ensemble by solving the HF equations for the target nucleus in the equal-filling approximation where occupation numbers of $m$-substates of partially filled orbitals are set to equal values between $0$ and~$1$.
The irreducible density matrices of this ensemble vanish by construction \cite{Stroberg:2017,Stroberg:2019}.  

We use the IMSRG(2) evolution to decouple the configurations \eqref{eq:def_configurations} from states that involve excitations of the core and from states containing nucleons in nonvalence particle states. 
The ensemble reference has trivial many-body correlations, so we can work in the quasi single-reference limit with noninteger occupation numbers. We achieve the desired decoupling using the arctangent generator for $\eta(s)$ \cite{Stroberg:2019} in two steps: First, we decouple the core from all excitations, just like in a ground-state calculation.
In a subsequent step, we additionally decouple the valence space from all excitations to the excluded space.
The final  valence-space Hamiltonian after re-normal-ordering with respect to the core%
\footnote{The re-normal-ordering step is not strictly necessary, but it makes the resulting interaction compatible with standard shell-model codes.}
can be written as
\begin{equation}
 \overline{H} =  E + \sum_{v_i v_j}f^{v_i}_{v_j}\nord{\AO^{v_i}_{v_j}} + \frac{1}{4}\sum_{\substack{v_i v_j \\ v_k v_l}}
  \Gamma^{v_iv_j}_{v_kv_l}\nord{\AO^{v_iv_j}_{v_kv_l}},
\end{equation}
where the explicitly shown terms are the core energy, single-particle
energies, and two-body matrix elements that are used as input for 
a shell-model diagonalization. The solutions of that diagonalization 
are given by
\begin{equation}
  \ket{\overline{\Psi}_n} = \sum_{v_1,\ldots,v_{A_v}} C^{(n)}_{v_1\ldots v_{A_v}}\aaO_{v_1}\ldots\aaO_{v_{A_v}}\ket{\Phi_{A_c}}\,,
\end{equation}
and they are related to the eigenstates of the initial Hamiltonian (up to 
truncation errors) by
\begin{equation}
  \ket{\Psi_n} = \UUO(\infty)\ket{\overline{\Psi}_n}\,.
\end{equation}
More details can be found in a recent review paper~\cite{Stroberg:2019}.

  \subsection{The nuclear matrix elements of \texorpdfstring{$0\nu\beta\beta$}{0νββ} decay}
  
  \subsubsection{Bare transition operator}
  Based on the light-Majorana neutrino-exchange mechanism, we can derive the transition operator  which in long-wavelength approximation for the outgoing electrons and without the recoil effect of the final nucleus,  is given by  \cite{Doi:1985,Bilenky:1987,Tomoda:1991,Simkovic99,Simkovic:2008,Stefanik:2015,Meng:2017}
\begin{align}
  O^{0\nu} 
   &=\frac{4\pi R}{g_A^2(0)} \iint \dd^3 r_1 \dd^3 r_2 \int\frac{\dd^3 q}{(2\pi)^3}\frac{e^{i \vec{q}\cdot \vec{r}_{12}}}{q}\notag\\
  &\hphantom{{}={}}\times \sum_m \dfrac{\mathcal {J}_{\mu}^\dagger(\vec{r}_1)\ket{m}\bra{m}  \mathcal {J}^{\mu\dagger}(\vec{r}_2)}{q+E_m-(E_I+E_F)/2},
\label{operator}
\end{align}
where $R=R_0 A^{1/3}$ is introduced to make the NME dimensionless, $\vec{r}_{12}=\vec{r}_1-\vec{r}_2$, and $q$ is the momentum transferred from leptons to hadrons.
The closure approximation is adopted, namely, the excitation energies of all possible intermediate states $\ket{m}$ are replaced by an estimate ``average" value $E_d=\langle E_m\rangle-(E_I+E_F)/2$ and the summation over these states can be eliminated by making use of the relation $\sum_m \ket{m}\bra{m}=1$, where $E_I$, $E_F$ are the energies of the initial and final nuclear state, respectively.
As a consequence, the summation of products of one-body matrix elements is approximated as one simple two-body matrix element.
An empirical formula value $E_d=\SI[parse-numbers=false,number-math-rm=]{1.12 A^{1/2}}{\MeV}$ proposed by \citet{Haxton1984PPNP} is adopted for the average excitation energy.
Since the average value of $q$ is around $1/\langle r_{12}\rangle\sim \SI{100}{\MeV}$, which is much larger than the average nuclear excitation energy ($10-20$ MeV) for long-ranged two nucleon processes,
the closure approximation is accurate at the $\SI{10}{\percent}-\SI{20}{\percent}$ level, as demonstrated  in Refs.~\cite{Pantis:1990,Faessler:1991,Senkov2013}.

The one-body charge-changing nucleon current operator $\mathcal {J}_{\mu}^\dagger(\vec{r})$  is employed,
\begin{align}
  \mathcal J_{\mu}^\dagger (\vec{r})
  &=\bar \psi (\vec{r})
  \Biggl[g_V(\vec{q}^2)\gamma_\mu - g_A(\vec{q}^2)\gamma_\mu \gamma_5 \notag\\
  &\hphantom{{}={}} -  ig_M(\vec{q}^2)\frac{\sigma_{\mu\nu}}{2m_p}q^\nu+g_P (\vec{q}^2)q_\mu \gamma_5
  \Biggr]\tau^+\psi(\vec{r}),
\label{Ncurrent}
\end{align}
where $\psi$ is the nucleon field operator and $m_p$ is the proton mass, $\tau^+$ is an isospin raising operator with the nonzero matrix element $\braket{p|\tau^+|n}=1$, and $\sigma_{\mu\nu}=\frac{i}{2} [\gamma_\mu,\gamma_\nu]$ with $\gamma_i$ being a four-component Dirac matrix. The higher-order (N$^2$LO in the power counting of the chiral EFT) corrections  to the one-body current operator are taken into account using the momentum-dependent form factors $g_V(\vec{q}^2)$, $g_A(\vec{q}^2)$, $g_M(\vec{q}^2),$ and $g_P(\vec{q}^2)$, which in the zero-momentum-transfer limit are the vector, axial-vector, weak-magnetism, and induced pseudoscalar coupling constants, respectively. The single-nucleon form factors are chosen as
\bsub\begin{align}
g_V(\vec{q}^2) &= g_V(0)\left(1+ \vec{q}^2/\Lambda^2_V\right)^{-2},\\
g_A(\vec{q}^2) &= g_A(0)\left(1+ \vec{q}^2/\Lambda^2_A\right)^{-2},\\
g_M(\vec{q}^2) &= g_V(\vec{q}^2)\left(1+\kappa_1\right),\\
g_P(\vec{q}^2) &= g_A(\vec{q}^2) \left(\dfrac{2m_p}{ \vec{q}^2+m^2_\pi}\right),
\end{align}
\esub
where $g_V(0)=1$, $g_A(0)=1.27$, the anomalous nucleon iso\-vector magnetic moment is $\kappa_1=\mu^{(a)}_n-\mu^{(a)}_p=3.7$, and the cutoff values are $\Lambda_V=\SI{0.85}{\GeV}$ and 
$\Lambda_A=\SI{1.09}{\GeV}$.

 The nonrelativistic reduction form of the transition operator $O^{0\nu}$ in \eqref{operator} is adopted in the present calculations. Previous studies have shown that relativistic corrections are approximately 5\% for this operator \cite{Song:2014,Yao:2015}.   The transition operator is recast into three parts: Fermi (F), Gamow-Teller (GT),  and tensor (T),
\begin{align}
 O^{0\nu} 
 &=   O^{0\nu}_F +  O^{0\nu}_{GT} +   O^{0\nu}_T\nonumber\\
 &=    H^{0\nu}_{F,0}(r_{12})\tau^+_1\tau^+_2
 +  H^{0\nu}_{GT,0 }(r_{12}) \vec{\sigma}_1\cdot\vec{\sigma}_2\tau^+_1\tau^+_2 \nonumber\\
&\hphantom{{}={}} + H^{0\nu}_{T,2}(r_{12}) \left[3(\vec{\sigma}_1\cdot \hat{\vec{r}}_{12})(\vec{\sigma}_2\cdot \hat{\vec{r}}_{12})
 -\vec{\sigma}_1\cdot\vec{\sigma}_2 \right]\tau^+_1\tau^+_2
\end{align}
with $\hat{\vec{r}}_{12}=\vec{r}_{12}/\lvert\vec{r}_{12}\rvert$. The neutrino potentials are given by
\begin{align}
 H^{0\nu}_{\alpha,L}(r_{12})&=
\dfrac{2R}{\pi g^2_A(0)} \int^\infty_0 \dd q \, q^2 \dfrac{ h_\alpha (\bm q^2)}{q(q+E_d)} j_L(qr_{12}), 
\end{align}
where $\alpha$ is the index for F, GT, or T, respectively. The function $j_L(qr_{12})$ is the spherical Bessel function of rank $L$.
The operators $h_{\alpha}$ are
\beq
h_{F}(\vec{q}^2) = -g^2_V(\vec{q}^2)
\eeq
for the Fermi part ($L=0$),
\begin{align}
h_{GT}(\vec{q}^2) &= g^2_A(\vec{q}^2)-\dfrac{2}{3} \dfrac{\vec{q}^2}{2m_p} g_A(\vec{q}^2)g_P(\vec{q}^2)\nonumber\\
&\hphantom{{}={}}+\dfrac{1}{3} \dfrac{\vec{q}^4}{4m^2_p} g^2_P(\vec{q}^2) +
\dfrac{2}{3} \dfrac{\vec{q}^2}{4m^2_p} g^2_M(\vec{q}^2) 
\end{align}
for the GT part ($L=0$), and
\beq
 h_{T}(\vec{q}^2) =  \dfrac{2}{3} \dfrac{\vec{q}^2}{2m_p} g_A(\vec{q}^2)g_P(\vec{q}^2)
-\dfrac{1}{3} \dfrac{\vec{q}^4}{4m^2_p} g^2_P(\vec{q}^2) +
\dfrac{1}{3} \dfrac{\vec{q}^2}{4m^2_p} g^2_M(\vec{q}^2)
\eeq
for the tensor part ($L=2$).

The above transition operator is \emph{essentially}%
\footnote{The value $g_A(0)=1.254$, and $g_M(\vec{q}^2)=\kappa_1 g_V(\vec{q}^2)$ were used in Ref.~\cite{Simkovic99}. Besides, an additional factor $1-m^2_\pi/\Lambda^2_A$ was multiplied onto $g_P(\vec{q}^2)$. Different from Ref.\cite{Deppisch:2020ztt},  we didn't find any sign problem for the tensor terms. The change of the relative sign between Gamow-Teller ($L=0$) and tensor ($L=2$) terms in Ref. \cite{Simkovic99} must be due to the factor of $i^L$ in the plane wave expansion  $e^{i \bm{q}\cdot \bm{r}_{12}}=4\pi\sum_{LM}i^L j_L(qr_{12})Y^\ast_{LM}(\hat{\bm{q}})Y_{LM}(\hat{\bm{r}}_{12})$, where $Y_{LM}$ is the spherical harmonics.}
the same as that  in Ref.~\cite{Simkovic99} and consistent with the operator \cite{Cirigliano:2018PRC} derived from chiral EFT based on the light-Majorana neutrino-exchange mechanism, which was adopted in the recent \emph{ab initio} QMC calculation \cite{Pastore:2018}, except for their choice of $E_d=0$ in (\ref{operator}).
We neglect the newly discovered LO contact operator \cite{Cirigliano:2018}, the extra parts of N$^2$LO corrections \cite{Cirigliano:2018PRC} that cannot be absorbed into parametrizations of the single-nucleon form factors, and the two-body weak currents \cite{Menendez:2011,Wang2018} that appear at N$^3$LO in the chiral EFT. Even though our adopted transition operator is not derived consistently with the chiral interaction, it does not change the conclusions that we will draw from our benchmark calculations because the same strong and weak operators are used in all the many-body calculations.
In addition, this facilitates a direct comparison with the results of Ref.~\cite{Novario2020}.

  \subsubsection{IMSRG(2)-evolved transition operator}
 Rewriting the two-body charge-changing transition operator $O^{0\nu}$ in second-quantized form,  
 \beq
  O^{0\nu} = \dfrac{1}{4}\sum_{pp'nn'} O^{pp'}_{nn'} \nord{A^{pp'}_{nn'}}
 \eeq
 where $p, p'$ and $n, n'$ are indices for protons and neutrons, respectively, we generate the evolved  operator  $ O^{0\nu}(s)  =e^{ \Omega(s)}  O^{0\nu}  e^{-\Omega(s)}$  using the BCH formula truncated at the NO2B level, 
\beqn
 O^{0\nu}(s)  
 &=&  O^{0\nu} + [ \Omega(s),  O^{0\nu}] + \dfrac{1}{2!}[ \Omega(s), [ \Omega(s),  O^{0\nu}]] + \cdots \nonumber\\
 &\equiv&  \dfrac{1}{4}\sum_{pp'nn'} \overline{O}^{pp'}_{nn'} \nord{A^{pp'}_{nn'}}.
\eeqn
For convenience, we introduce the two-body operator $D^{(2)}$ for the commutator under the NO2B approximation,
\begin{align}
D^{(2)}
&=[\Omega,   O^{0\nu}]^{(2)}\nonumber\\
&= [ \Omega^{(1)},  O^{0\nu}]^{(2)} + [ \Omega^{(2)},  O^{0\nu}]^{(2)}\nonumber\\
&\equiv \dfrac{1}{4}\sum_{pp'nn'} \left( D^{pp'}_{nn'}(\text{1B})+D^{pp'}_{nn'}(\text{2B})
\right) \nord{A^{pp'}_{nn'}}.
\label{CommutatorDBD}
\end{align}
The contributions involving the one-body $ \Omega^{(1)}$ and two-body $ \Omega^{(2)}$ parts of $\Omega$ in natural-orbital basis (i.e., $\lambda^p_q=n_p \delta^p_q$)  are given by
\begin{align}
D^{pp'}_{nn'}(\text{1B})
&= \sum_{p_1}\left(\Omega^{p}_{p_1} O^{p_1p'}_{nn'}+\Omega^{p'}_{p_1} O^{pp_1}_{nn'}\right) \nonumber\\
&\hphantom{{}={}}-\sum_{n_1}\left(\Omega^{n_1}_{n} O^{pp'}_{n_1n'}+\Omega^{n_1}_{n'} O^{pp'}_{nn_1}\right)
\,,
\label{eq:1bpart}
\shortintertext{and} 
D^{pp'}_{nn'}(\text{2B})
 &=  \dfrac{1}{2}\sum_{p_1p_2} \Omega^{pp'}_{p_1p_2}   O^{p_1p_2}_{nn'}  (1-n_{p_1}-n_{p_2})\nonumber\\
&\hphantom{{}={}}-\dfrac{1}{2}\sum_{n_1n_2} O^{pp'}_{n_1n_2} \Omega^{n_1n_2}_{nn'}
(1-n_{n_1}-n_{n_2}) \,,\nonumber\\
&\hphantom{{}={}}+
  \sum_{p_1n_1} (n_{p_1}-n_{n_1})  \left[
    \Omega^{n_1p'}_{n'p_1}  O^{p_1p}_{n_1n}
   -\Omega^{n_1p}_{n'p_1}   O^{p_1p'}_{n_1n}\right.
  \nonumber\\
&\hphantom{{}={}}\left.\qquad
{}+ \Omega^{n_1p}_{np_1}   O^{p_1p'}_{n_1n'}
- \Omega^{n_1p'}_{np_1}   O^{p_1p}_{n_1n'}  \right].
\label{eq:trans_op_2B}
\end{align}
The predominant contribution to $D^{(2)}$ is the two-body  part $D^{pp'}_{nn'}(\text{2B})$, which can be divided into  particle-particle ($pp$) and hole-hole ($hh$) terms corresponding to the first two terms in \eqref{eq:trans_op_2B}.
The last term is the particle-hole ($ph$) term.
As shown in Sec.~\ref{results}, the combination of $pp$ and $hh$ contributions enhances the two-body transition matrix element $O^{pp'}_{nn'}$, while the $ph$ contribution quenches the matrix element.

Since $\Omega(s)$ conserves charge, no zero- or one-body terms are generated when we evaluate the commutator \eqref{CommutatorDBD} (induced higher-body operators are truncated), and the resulting operator has the same isospin structure as the initial transition
operator itself.  This means that we can use Eqs.~\eqref{CommutatorDBD}--\eqref{eq:trans_op_2B} to recursively evaluate the BCH series by replacing $O^{0\nu}$ with the appropriate nested commutator $[\Omega,O^{0\nu}]^{(n)}$. 

The renormalization effect on $O^{0\nu}$ depends on the choice of the reference state $\ket{\Phi}$, which controls the correlations built into the normal-ordered basis operators and the occupation numbers, as well as the generator $\eta$, which determines the unitary transformation operator $e^{\Omega}$:

\begin{itemize}
    \item  In the VS-IMSRG, we choose the reference state as a spherically-symmetric uncorrelated ensemble of multiple Slater determinants. For the calculations presented here, we use a reference with the $N$ and $Z$ of the parent nucleus. In contrast, the MR-IMSRG or IM-GCM calculation uses an ensemble of two \emph{correlated} states as the reference state which has nonvanishing irreducible density matrices entering into the normal ordering. 

    \item In the VS-IMSRG, the unitary transformation operator $e^{\Omega}$ decouples the valence-space ($v$) states from both the core ($c$) and nonvalence particle ($q$) states. 
    The nonzero matrix elements of $\eta$ (and therefore the largest matrix elements of $\Omega$) are those leading to excitations out of the core or valence orbits. 
    In contrast, the MR-IMSRG decouples the reference state from its multiparticle-multihole ($np$-$nh$) excitations. The resulting  $\Omega$ is dominated by the matrix elements that are connecting the reference state with low-energy excited states. 
    
\end{itemize}

In an exact calculation with untruncated operators, these differences must be compensated by the changes in the transformed Hamiltonian and nuclear wavefunctions. As shown below, the NMEs for $0\nu\beta\beta$ decays by the two approaches 
in general agree with each other with differences at the \SI{10}{\percent} level,
consistent with our expectation. Some cases show relatively larger differences that can be attributed to the adopted NO2B approximation (see \hyperref[appendix]{appendix}).
The spread of the results by the IMSRG(2)-based approaches and the IT-NCSM  gives an estimate of the uncertainty due to these approximations.

   \subsubsection{Nuclear matrix elements}
   
The NME of ground-state to ground-state $0\nu\beta\beta$ transition is given by
\beq\label{eq:me}
M^{0\nu} = \braket{ \Psi_F | \overline{O}^{0\nu} | \Psi_I} \,,
\eeq
where $\ket{ \Psi_{I/F}} $ is the ground-state wavefunction of the initial and final nuclei, respectively. The NME is usually decomposed into the contributions from the decaying pair of neutrons coupled to a given angular momentum $J$,
\beqn
\label{eq:MJ}
M^{0\nu} 
&=&\sum_J M^{0\nu}_J
\eeqn
where the $J$-component $M^{0\nu}_J$ is determined by
\begin{align}
M^{0\nu}_J
&=\sum_{p\leq p'; n\leq n'}\dfrac{(2J+1)}{\sqrt{(1+\delta_{pp'})(1+\delta_{nn'})}} 
 \tensor*[^{J}]{\overline{\mathbb{O}}}{^{pp'}_{nn'}}
 \tensor*[^{J}]{\rho}{^{pp'}_{nn'}},
\end{align}
with $\tensor*[^{J}]{\overline{\mathbb{O}}}{^{pp'}_{nn'}}$ being the  \emph{normalized} two-body  matrix element of the IMSRG-evolved transition operator, which is related to the \emph{unnormalized} two-body matrix element $\tensor*[^{J}]{\overline{O}}{^{pp'}_{nn'}}$ by
\begin{align}
\tensor*[^J]{\overline{\mathbb{O}}}{^{pp'}_{nn'}} 
 &= \dfrac{\tensor*[^J]{\overline{O}}{^{pp'}_{nn'}}}{\sqrt{(1+\delta_{pp'})(1+\delta_{nn'})}}\nonumber\\ 
 &=\dfrac{1}{\sqrt{(1+\delta_{pp'})(1+\delta_{nn'})}} \sum_{\substack{m_pm_{p'} \\ m_nm_{n'}}}\nonumber\\ 
 &\hphantom{{}={}}\times\braket{j_p m_p j_{p'}m_{p'}|JM}
  \braket{j_n m_n j_{n'}m_{n'}|JM} \overline{O}^{pp'}_{nn'}
 \end{align}
 and the two-body transition density
 \beqn
 \tensor*[^{J}]{\rho}{^{pp'}_{nn'}}
 = -\dfrac{1}{\sqrt{2J+1}} \braket{\Psi_F | [a^\dagger_{p}a^\dagger_{p'}]^J
 [\tilde a_{n} \tilde a_{n'}]^J| \Psi_I}.
 \eeqn
 The square brackets $[\cdots]^J$ mean a coupling of two spherical tensors to angular momentum $J$. We note that in the IT-NCSM calculation the bare transition operator $O^{0\nu}$ is used instead of the evolved operator $\overline{O}^{0\nu}$.
   
   \section{Results and discussion}
   \label{results}
    
\begin{figure}
\centering 
\includegraphics[width=8.4cm]{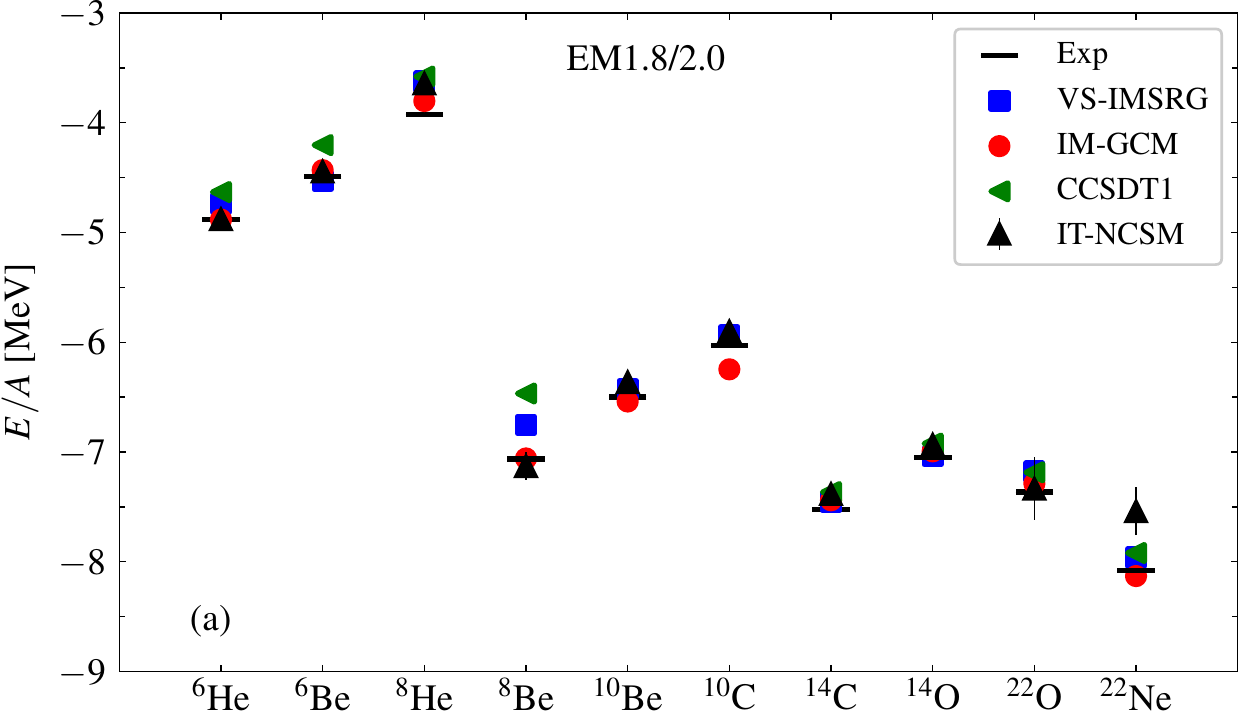}
\includegraphics[width=8.4cm]{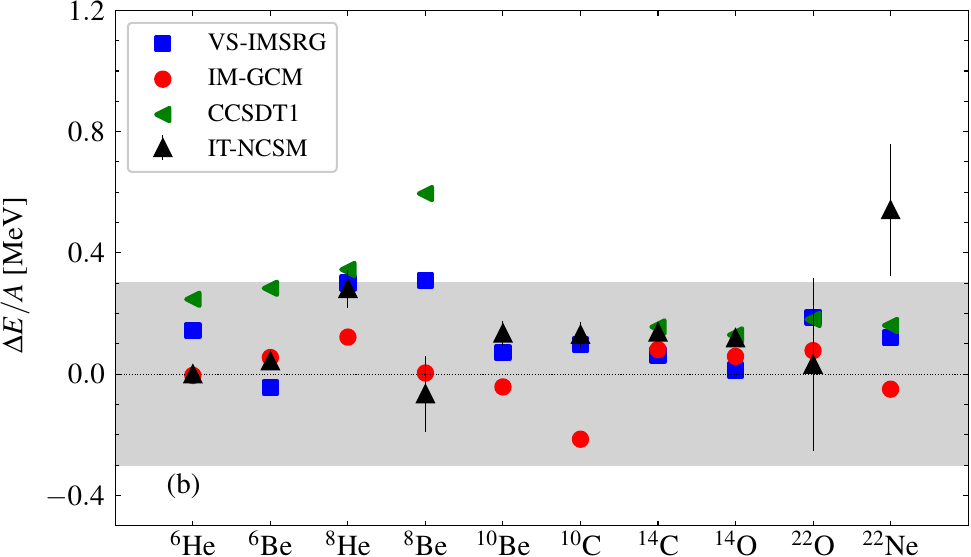} 
\caption{\label{fig:energy}(Color online)  (a) Energies per nucleon ($E/A$) of the light nuclei from the VS-IMSRG, IM-GCM, and IT-NCSM calculations,  in comparison with those from the CCSDT1 calculations \cite{Novario2020} and with data \cite{NNDC}. (b) The discrepancy  of the $E/A$ between each model calculation and corresponding data. The gray band indicates the energy spread within \SI{0.3}{\MeV} per nucleon. The error bars of IT-NCSM are for the $\NMax$ extrapolation uncertainties.}
\end{figure}

  Before showing the NMEs of $0\nu\beta\beta$ decays, we compare the ground-state energies of  initial and final nuclei from different calculations in Fig.~\ref{fig:energy}.
  The energies are reasonably reproduced in all the calculations with the discrepancy within \SI{0.3}{\MeV} per nucleon, except for $\nuclide[8]{Be}$.
  The ground state of $^{8}$Be is expected to be dominated by a two-$\alpha$ cluster structure, the description of which requires configurations of four-particle-four-hole ($4p$-$4h$) or even higher excitations \cite{Myo:2014} and is challenging for the CCSDT1 and VS-IMSRG(2), which underestimate the energy by about \SI{8}{\percent} and \SI{4}{\percent}, respectively.
  In contrast, the IM-GCM works rather well for $^{8}$Be as the configurations of $np$-$nh$ excitations are relatively easier included by mixing intrinsically large deformed mean-field states explicitly. For the $sd$-shell nuclei $^{22}$O and $^{22}$Ne, converging IT-NCSM calculations is challenging \cite{Hergert:2013ij}.  The ground state wavefunction of \nuclide[22]{Ne} in the IT-NCSM calculation at $\NMax=8$, the largest model space considered, has a $J$ expectation value of \num{0.9}. This indicates admixture of low-lying excited states (an artifact of the importance truncation) that could be the cause of the difference between the IT-NCSM and the other results.

 \begin{figure}
\centering
\includegraphics[width=8.4cm]{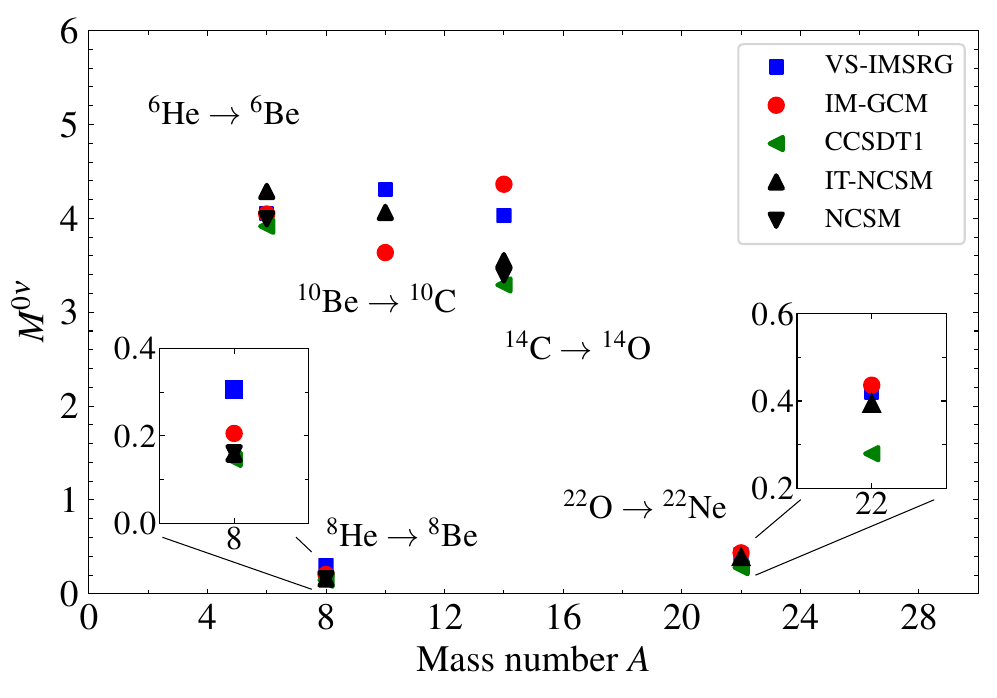} 
\caption{\label{fig:NMEs}(Color online) The $0\nu\beta\beta$-decay NMEs of the light nuclei from the VS-IMSRG, IM-GCM, and IT-NCSM calculations, in comparison with the  results of NCSM and CCSDT1 calculations from Refs.~\cite{Novario2020,Novario2020:private}.  }
\end{figure}

\begin{figure}
  \centering
  \includegraphics{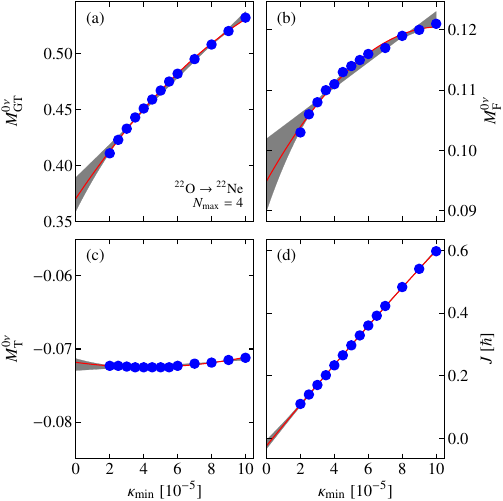}
  \caption{\label{fig:O2Ne22-kappa}%
    (Color online)
    Threshold extrapolation of (a-c) the NME for the $0\nu\beta\beta$ decay of \nuclide[22]{O} to \nuclide[22]{Ne}, and (d) the \nuclide[22]{Ne} ground-state total angular momentum.
    Finite-threshold results are marked by blue dots, the red line is the quadratic extrapolation polynomial, and the shaded band marks the range of different extrapolations.
    Note the different scales of the panels.
  }
\end{figure}

 \begin{figure}
\centering
\includegraphics[width=\columnwidth]{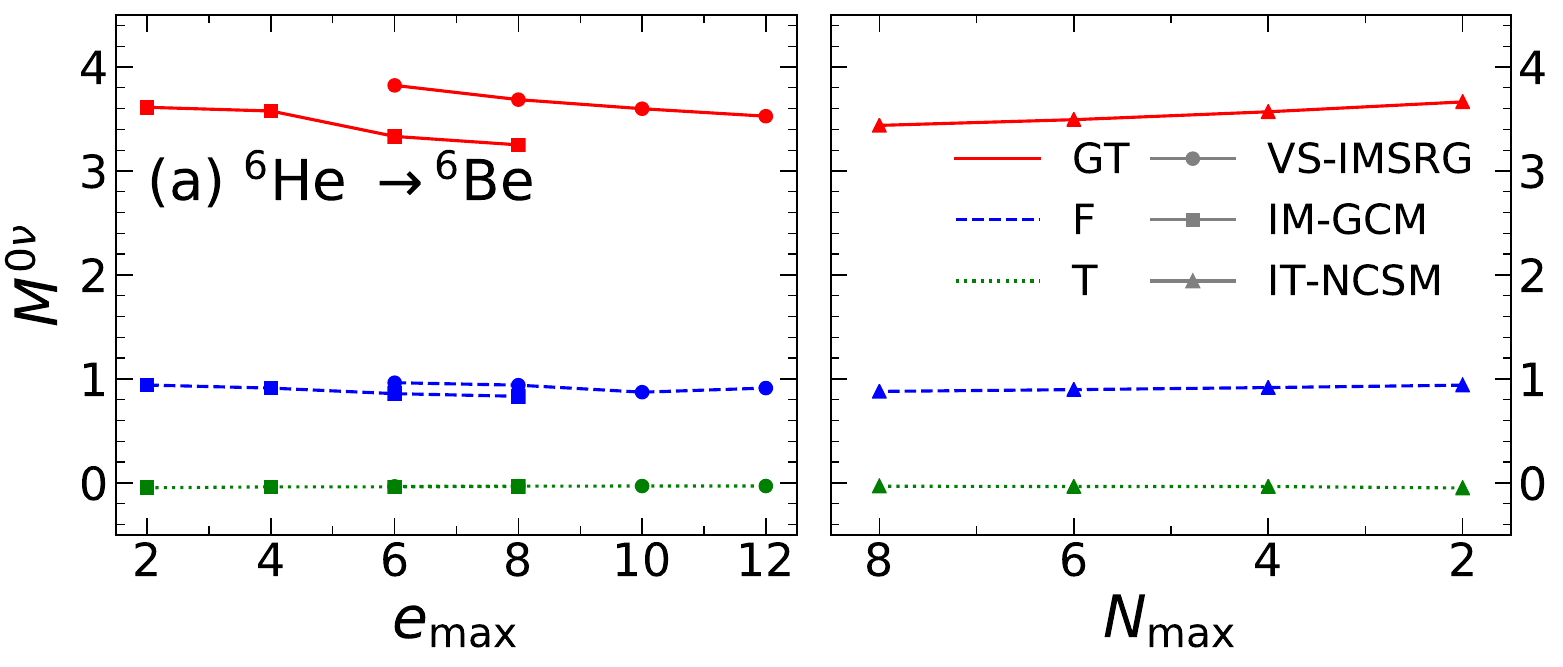} 
\includegraphics[width=\columnwidth]{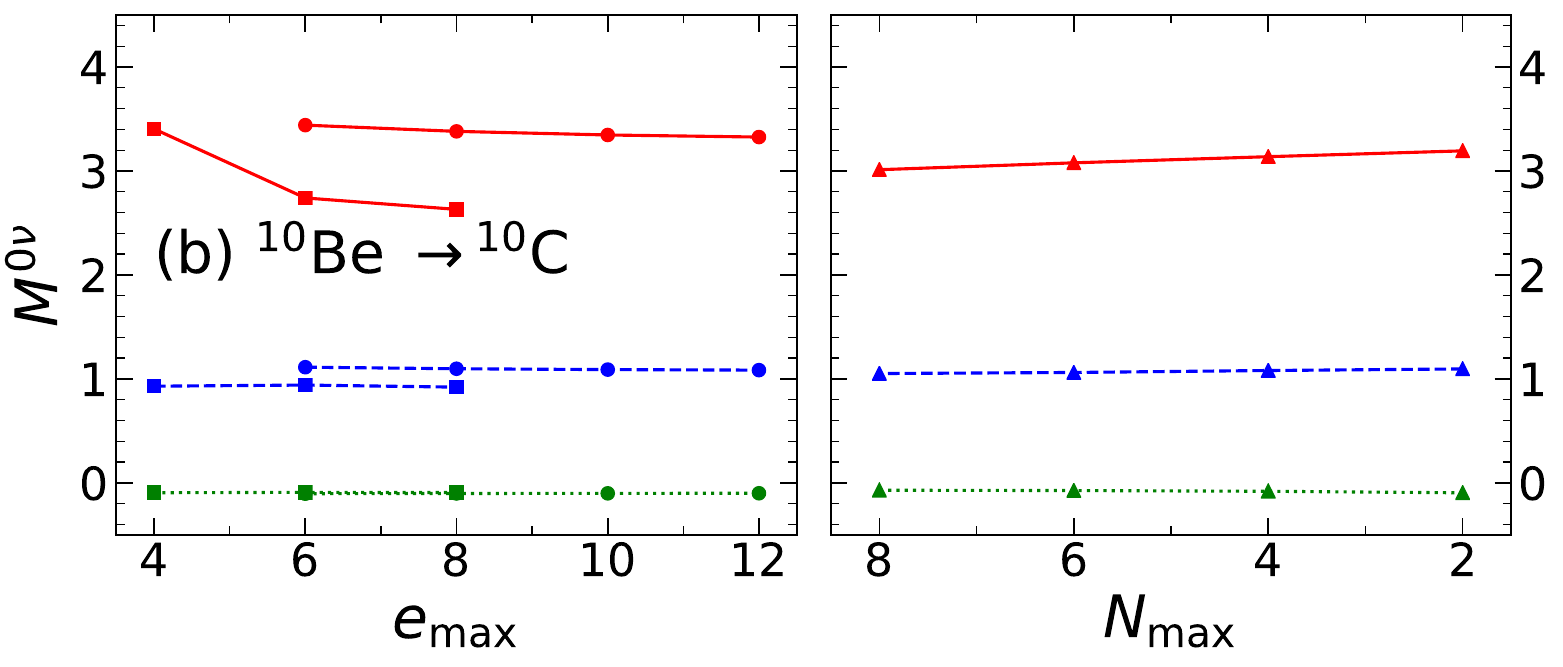}
\includegraphics[width=\columnwidth]{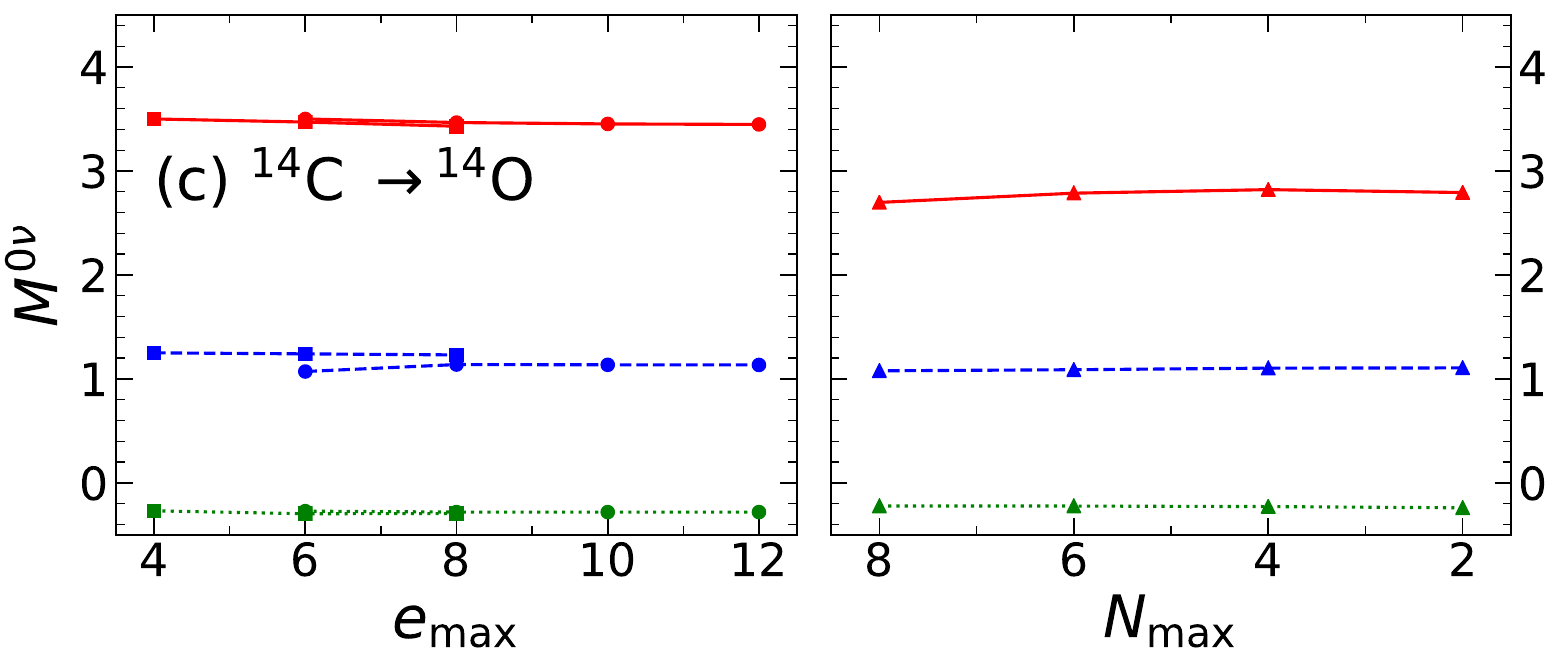} 
\caption{\label{fig:decompose_eMax1}(Color online)  The NMEs of the isospin-conserving $0\nu\beta\beta$ decays (a) $\nuclide[6]{He} \to \nuclide[6]{Be}$, (b) $\nuclide[10]{Be} \to \nuclide[10]{C}$, and (c) $\nuclide[14]{C} \to \nuclide[14]{O}$ from the VS-IMSRG, IM-GCM (left), and the IT-NCSM (right) as a function of $\eMax$ and $\NMax$, respectively.}
\end{figure}

 \begin{figure}
\centering
\includegraphics[width=\columnwidth]{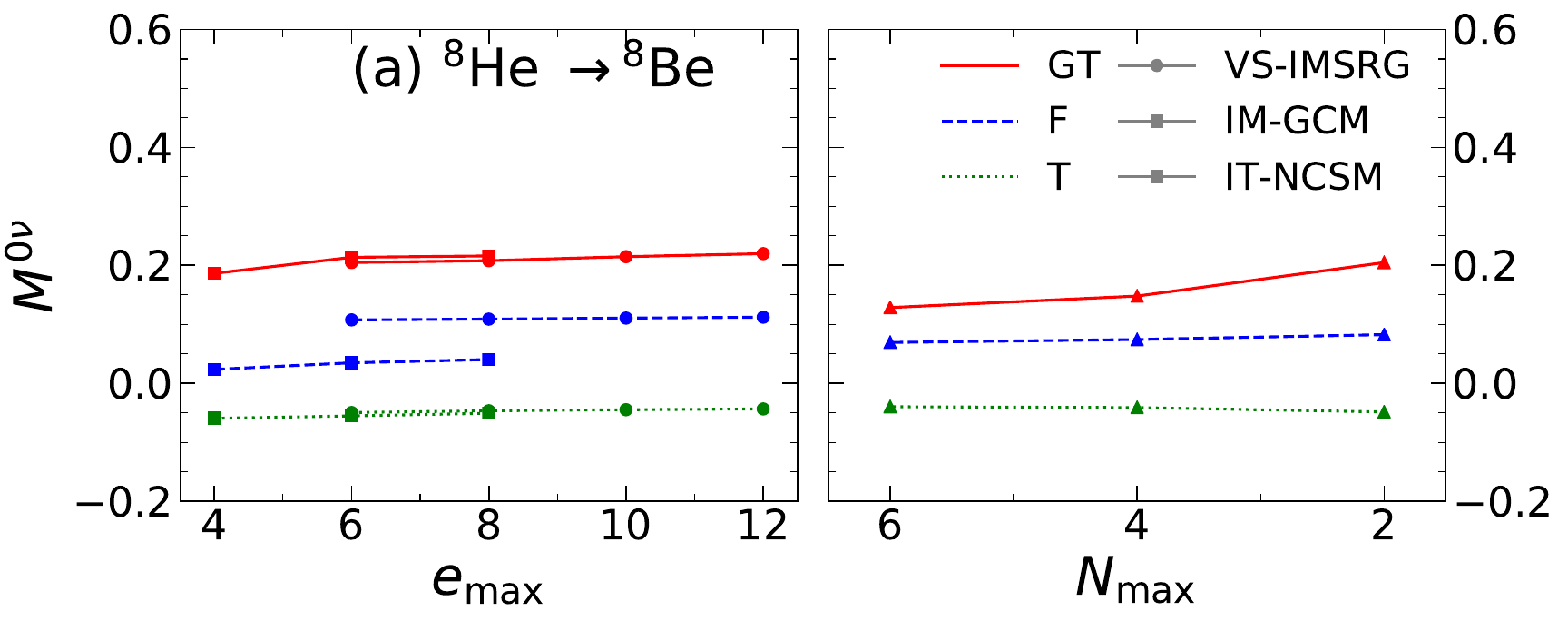} 
\includegraphics[width=\columnwidth]{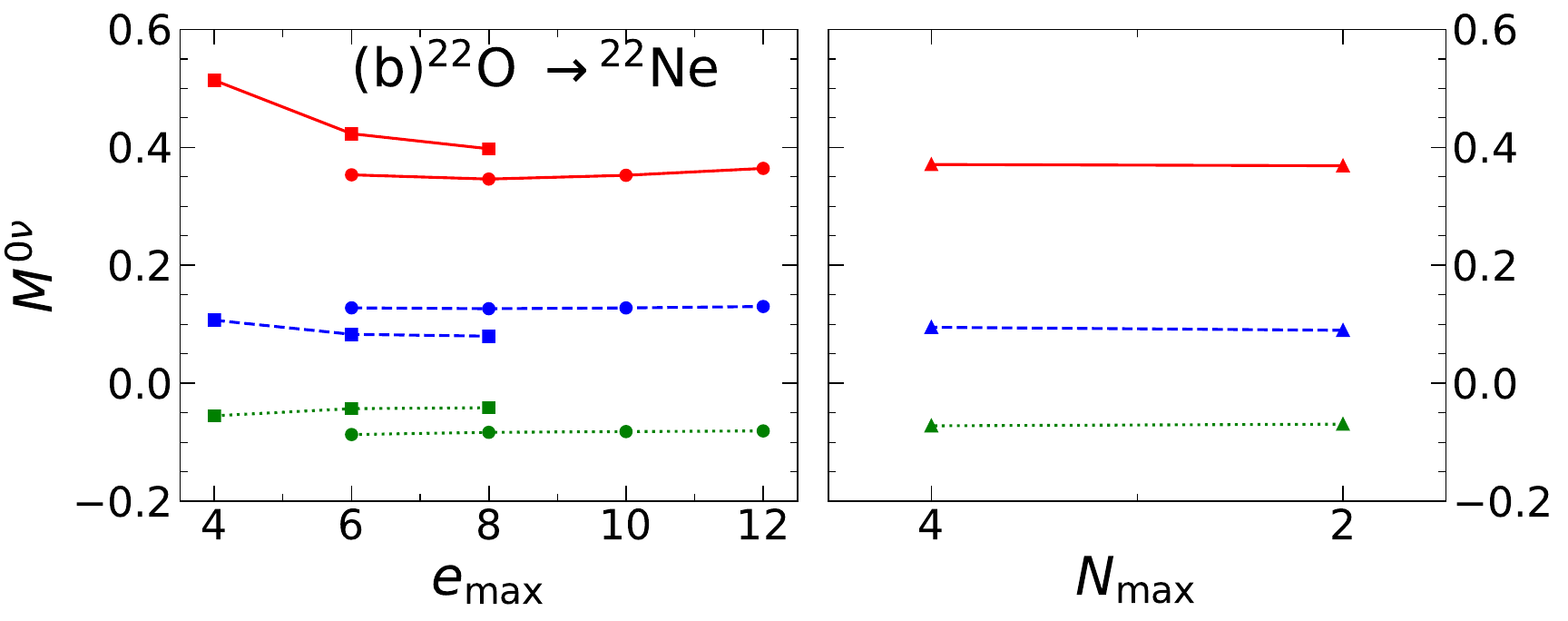} 
\caption{\label{fig:decompose_eMax2}%
(Color online) Same as Fig. \ref{fig:decompose_eMax1}, but for  the isospin-changing $0\nu\beta\beta$ decays (a) $\nuclide[8]{He} \to \nuclide[8]{Be}$ and (b) $\nuclide[22]{O} \to \nuclide[22]{Ne}$. }
\end{figure}

The total  NMEs of the $0\nu\beta\beta$ decays  are compared in Fig.~\ref{fig:NMEs}, where the NME of the transition  $\nuclide[22]{O} \to \nuclide[22]{Ne}$ by the IT-NCSM using $\NMax=4$ is taken for comparison as this calculation produces correct $J$ expectation values for the ground states of both nuclei. Similar to the results found in Ref.~\cite{Pastore:2018}, the NME of the isospin-conserving transition is around 4.0 (with the GT part around 3.0 and Fermi part around 1.0), while that of the isospin-changing transition is about an order of magnitude smaller.   
For the $\Delta T=0$ transitions, we explored the dependence on different choice of reference ensemble in the IM-GCM calculations, and found it to be around \SI{10}{\percent} for $\nuclide[14]{C}$, as indicated with an error bar.  Taking into account the uncertainty, the NME for $\nuclide[14]{C} \to \nuclide[14]{O}$ obtained with the two variants of IMSRG are consistent.
For $\nuclide[10]{Be}\to\nuclide[10]{C}$, reference dependence is around \SI{5}{\percent}.  The result of the IT-NCSM calculation is sandwiched between the VS-IMSRG and IM-GCM results with a discrepancy less than \SI{10}{\percent}. For $^{6}$He, this dependence is negligible.  For the $\Delta T=2$ transitions,  the NMEs of the CCSDT1 calculation \cite{Novario2020} starting from a deformed reference state (of final nucleus) are taken for comparison in Fig.~\ref{fig:NMEs}. It is seen that the VS-IMSRG(2) overestimates the NME for $\nuclide[8]{He}$ which might be due to the issue that the collective correlation of the two-$\alpha$ clustering structure in $\nuclide[8]{Be}$ is not well captured and this correlation could quench the NME significantly. It is also shown in Ref.~\cite{Novario2020} that the choice of the reference state to be  $\nuclide[8]{He}$ in the CCSDT1 would also significantly overestimate the NME (which is around 0.6). The reference-state dependence is shown again in the NME of $\nuclide[22]{O}$, which is predicted to be around 0.856 or 0.279 in the CCSDT1 calculation \cite{Novario2020:private} if the reference state is chosen as $\nuclide[22]{O}$ or $\nuclide[22]{Ne}$, respectively.  

The IT-NCSM calculation of the transition NME of $\nuclide[22]{O}\to\nuclide[22]{Ne}$  shows sizable dependence on the importance threshold $\kappa_{\text{min}}$, so we perform a threshold extrapolation using a quadratic polynomial, as shown in Fig.~\ref{fig:O2Ne22-kappa}(a-c).
The uncertainties are obtained by fitting linear and cubic polynomials to the full set, and by fitting quadratic polynomials
to everything except for the one or two lowest-$\kappa_{\text{min}}$ points. 
This gives 5 fits, and we take the differences from the minimum and maximum values to the value of the first quadratic fit as uncertainty.
The extrapolated NME at $\NMax=4$ is $0.394^{+0.026}_{-0.017}$. Panel (d) shows that the total angular momentum expectation value is also strongly dependent on the threshold, a sign of admixture of a low-lying state with nonzero angular momentum due to the importance truncation. Lowering the threshold rapidly reduces the expectation value, showing that the states separate as the truncation is relaxed towards the full model space.
For the larger $\NMax$ we have to employ a larger importance threshold $\kappa_{\text{min}}=\num{3e-5}$ (basis dimension \num{43e6} at $\NMax=8$) in order to keep the size of the calculation manageable.
Unfortunately, the more severe truncation leads to an incomplete separation of the ground state and the extrapolation gives a nonzero $J$. Decreasing the threshold to improve this exceeds the computational resources available for this study.
However, the NMEs depend weakly on $\NMax$ so we expect only small changes from the larger model spaces.

 Next, we compare  the NMEs of Fermi,  Gamow-Teller, and tensor parts as a function of the model space in VS-IMSRG, IM-GCM, and IT-NCSM in Fig.~\ref{fig:decompose_eMax1} and Fig.~\ref{fig:decompose_eMax2} for the $\Delta T=0$ and $\Delta T=2$ transition, respectively. As expected, the tensor contribution to the NME is generally small for the $\Delta T=0$ transitions (less than \SI{10}{\percent}). However, for the $\Delta T=2$ transition from \nuclide[8]{He} to \nuclide[8]{Be}, the tensor can contribute up to \SI{25}{\percent} to the total NME, as the NME itself is small. The variation of the NME with model space is mainly driven by the GT part. For $^6$He, the GT matrix element decreases by about \SI{7}{\percent} and \SI{3}{\percent} when the model space increases from $\eMax=4$ to $\eMax=6$ and from $\eMax=6$ to $\eMax=8$ respectively in the IM-GCM calculations. For $^{10}$Be, these numbers are \SI{24}{\percent} and \SI{4}{\percent}, respectively. The spreads in the NMEs of $\nuclide[10]{Be}\to\nuclide[10]{C}$, and $\nuclide[14]{C}\to\nuclide[14]{O}$ are primarily due to the GT contribution, for reasons discussed below. We note that the NME for $^{6}$He to $^{6}$Be is comparable to the value in Ref. \cite{Basili2020} in which only the NN part of the interaction is taken into account in the NCSM calculation of the nuclear wavefunctions. Specifically, the NCSM with NN only and $\NMax=8$ predicts $M^{0\nu}=4.304$ \cite{Basili2020}, and the value by the IT-NCSM, which includes the 3N interaction, is 4.287. This seems to indicate that the 3N interaction has a negligible impact on the predicted NME for the transition from $^{6}$He to $^{6}$Be.

 \begin{figure}
\centering 
\includegraphics[width=7cm]{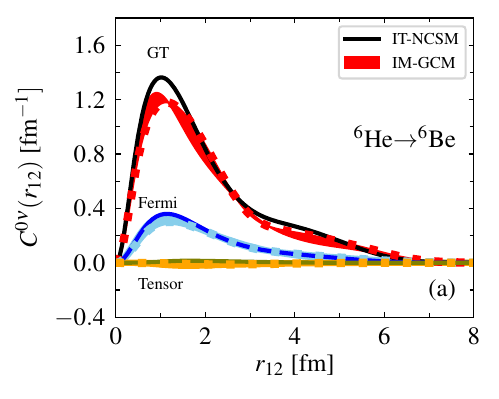}  
\includegraphics[width=7cm]{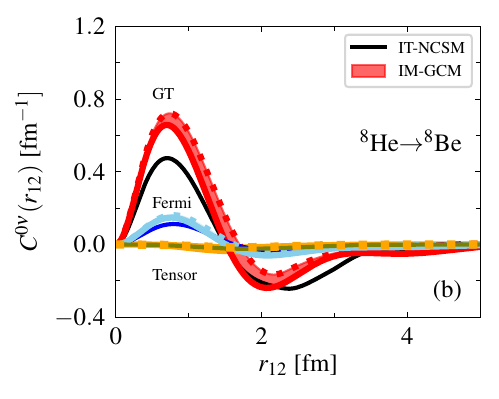}  
\caption{\label{fig:He2Be_TD}(Color online) The IMSRG evolution on the distribution $C^{0\nu}(r_{12})$ of the NME  as a function of the relative coordinate $r_{12}$ corresponding to the transitions (a)  from  $^{6}$He to $^{6}$Be and (b) from  $^{8}$He to $^{8}$Be from the IM-GCM calculation, in comparison with the result from the IT-NCSM calculations. The boundary of the shaded area indicated with dotted and solid curves  corresponds to the result using the bare and evolved transition operators, respectively. See text for details.}
\end{figure}

 \begin{figure}
\centering 
\includegraphics[width=7cm]{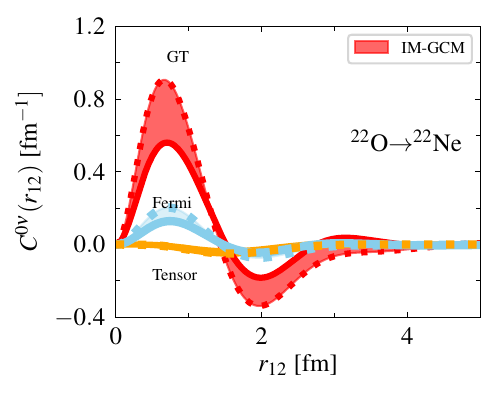}
\caption{\label{fig:O2Ne22_TD}(Color online) Same as Fig. \ref{fig:He2Be_TD}, but for the transition from $^{22}$O to $^{22}$Ne.
}
\end{figure}

 \begin{figure}
\centering 
\includegraphics[width=6.5cm]{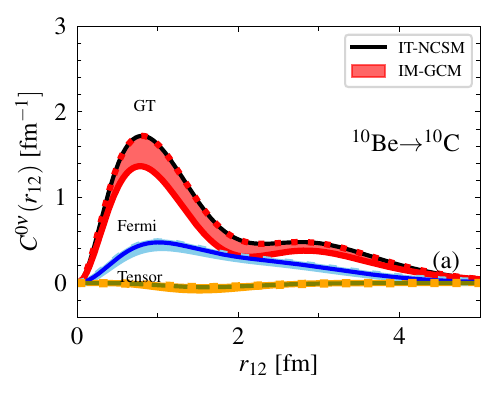}  
\includegraphics[width=6.5cm]{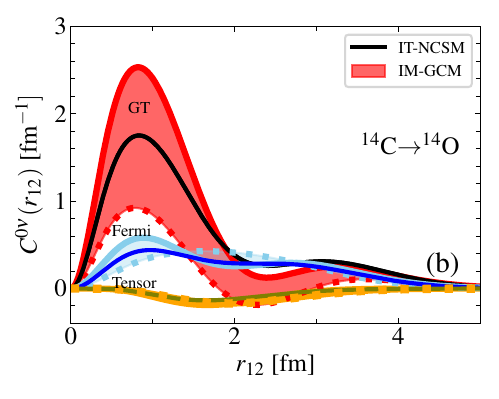} 
\caption{\label{fig:C2O14_TD}%
(Color online) Same as Fig. \ref{fig:He2Be_TD}, but for  $^{10}$Be and $^{14}$C.}
\end{figure}

We break down the NMEs further by introducing the transition distribution function $C^{0\nu}(r_{12})$, defined as 
\begin{equation}
    M^{0\nu} = \int^\infty_0 \dd r_{12}\, C^{0\nu}(r_{12}) \,,
\end{equation}
with $r_{12}=\lvert \vec{r}_1 - \vec{r}_2 \rvert$ being the relative distance between the two neutrons that are transformed into protons. Figure~\ref{fig:He2Be_TD} displays $C^{0\nu}(r_{12})$ for $^{6}$He and $^{8}$He from the IM-GCM and IT-NCSM calculations, where both the evolved and bare transition operators are used. The shaded area indicates the renormalization effect in the transition operator. We see that the shape of the functions from both calculations is similar, even though the height of the peak at short distance is somewhat different from each other. The differences between the two methods are most pronounced in the GT contributions.  It is worth mentioning that the transition distribution function $C^{0\nu}(r_{12})$ is scale- and scheme-dependent~\cite{Duguet:2015lq,More:2017eu,Gysbers:2019df} and any attempt at a quantitative comparison requires a careful accounting of the (IM)SRG transformations, even if both methods start from the same operators. Thus, a direct quantitative comparison of this distribution with the one obtained from QMC calculations with the AV18+IL7 interaction in Ref.~\cite{Pastore:2018} is not necessarily meaningful.  Qualitatively, we see similar global features like a single peak and the absence of nodes in the distribution $C^{0\nu}(r_{12})$ for $^{6}$He. In contrast, the  $C^{0\nu}(r_{12})$ for the isospin-changing transition from $^{8}$He to $^{8}$Be has a node around $r_{12}=\SI{1.5}{\fm}$. The cancellation between the long-range and the short-range contributions produces a small transition NME for $^{8}$He. A similar cancellation is also found in the GT part of the transition distribution function for $\nuclide[22]{O}$, as shown in Fig.~\ref{fig:O2Ne22_TD}, where the results of IT-NCSM are not plotted as it is challenging to extrapolate the distribution function in the similar way as what we have done in Fig.~\ref{fig:O2Ne22-kappa}. It is seen that the renormalization decreases the heights of peaks on both positive and negative sides, leading to a negligible effect on the total NME.

The distribution functions $C^{0\nu}(r_{12})$ for $^{10}$Be and  $^{14}$C are displayed in Fig.~\ref{fig:C2O14_TD}. One can clearly see that the impact of the IMSRG(2) evolution on the GT part of $\nuclide[10]{Be}$ and  $\nuclide[14]{C}$ is significantly larger than that on $\nuclide[6]{He}$ and  $\nuclide[8]{He}$. This effect is so strong that it leads to the underestimation of the GT transition matrix element for $^{10}$Be, and an overestimation for $^{14}$C. The size of the renormalization indicates that higher-order terms are more relevant for  these nuclei and their inclusion is expected to reduce the observed discrepancies. As discussed in the \hyperref[appendix]{appendix},  most of the direct contribution from the induced three-body  transition operator from the commutator $[\Omega^{(2)}, O^{0\nu}]^{(3)}$ to the NME has already been taken into account in the NO2B approximation. The residual part turns out to be negligible. However, its contribution to the commutator  $[\Omega^{(2)}, [\Omega^{(2)}, O^{0\nu}]^{(3)}]^{(2)}$ reduces the total \nuclide[14]{C} NME by about \SI{10}{\percent}. The low-lying states of spherical nuclei are generally dominated by noncollective configurations which are not included explicitly in the present IM-GCM calculation. The inclusion of noncollective configurations is expected to reduce further the discrepancies. Indeed, we find that the inclusion of neutron-proton isoscalar pairing fluctuations in the IM-GCM reduces the discrepancy by about \SI{5}{\percent}. With the corrections mentioned above, the discrepancy between the \nuclide[14]{C} NMEs by the IM-GCM and IT-NCSM  is within $\SI{10}{\percent}$.

 \begin{figure}
\centering
\includegraphics[width=8.6cm]{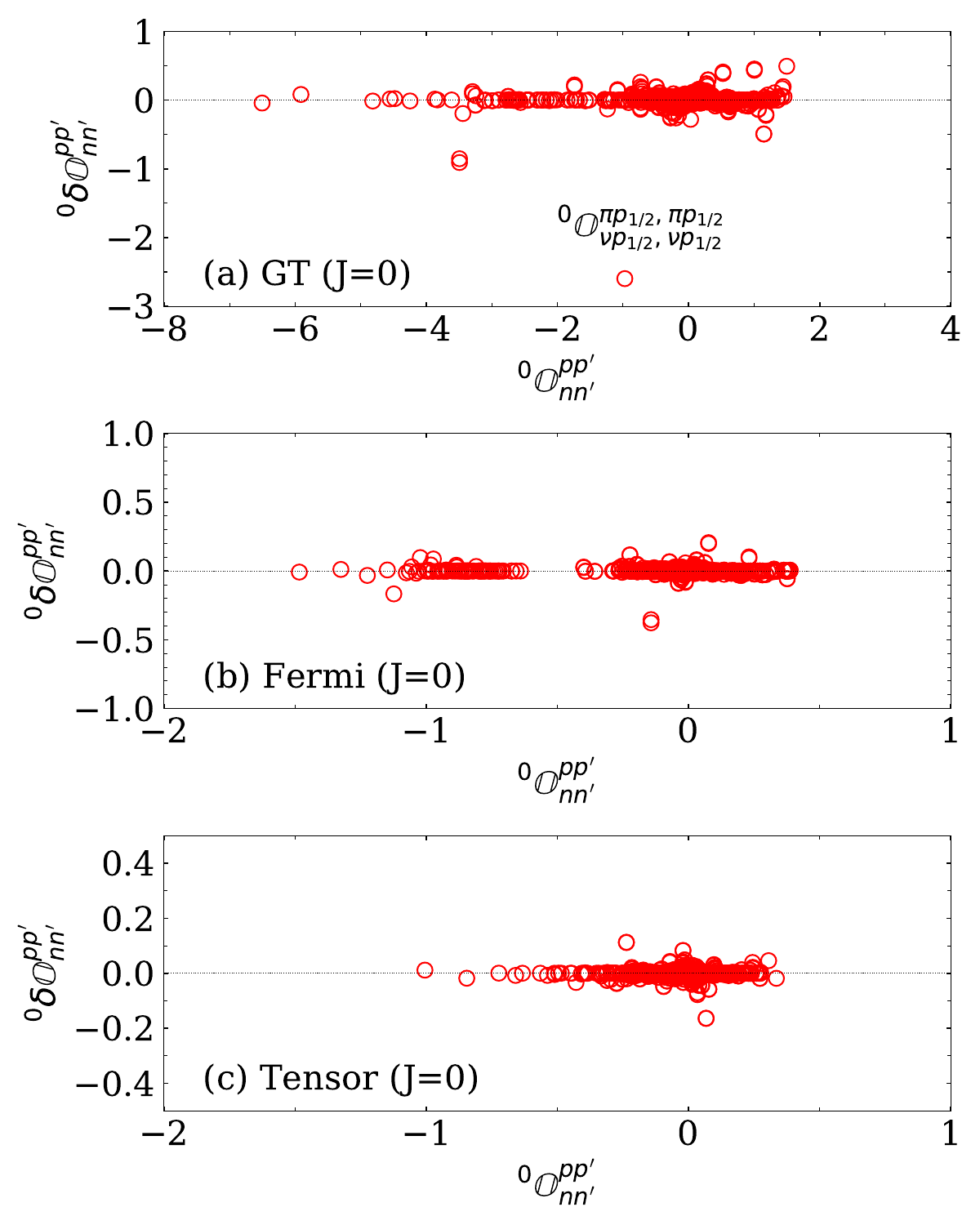}  
\caption{\label{fig:C2O14TBME}(Color online)  The renormalization effect on the \emph{normalized} two-body transition matrix element $\tensor*[^{J=0}]{\mathbb{O}}{^{pp'}_{nn'}}$ in the spherical HO basis with $\eMax=6$ and frequency $\hbar\Omega=\SI{16}{\MeV}$ in the IM-GCM calculation based on the  ensemble reference state of $^{14}$C and $^{14}$O. See text for further details. }
\end{figure}

 \begin{figure}
\centering 
\includegraphics[width=8.6cm]{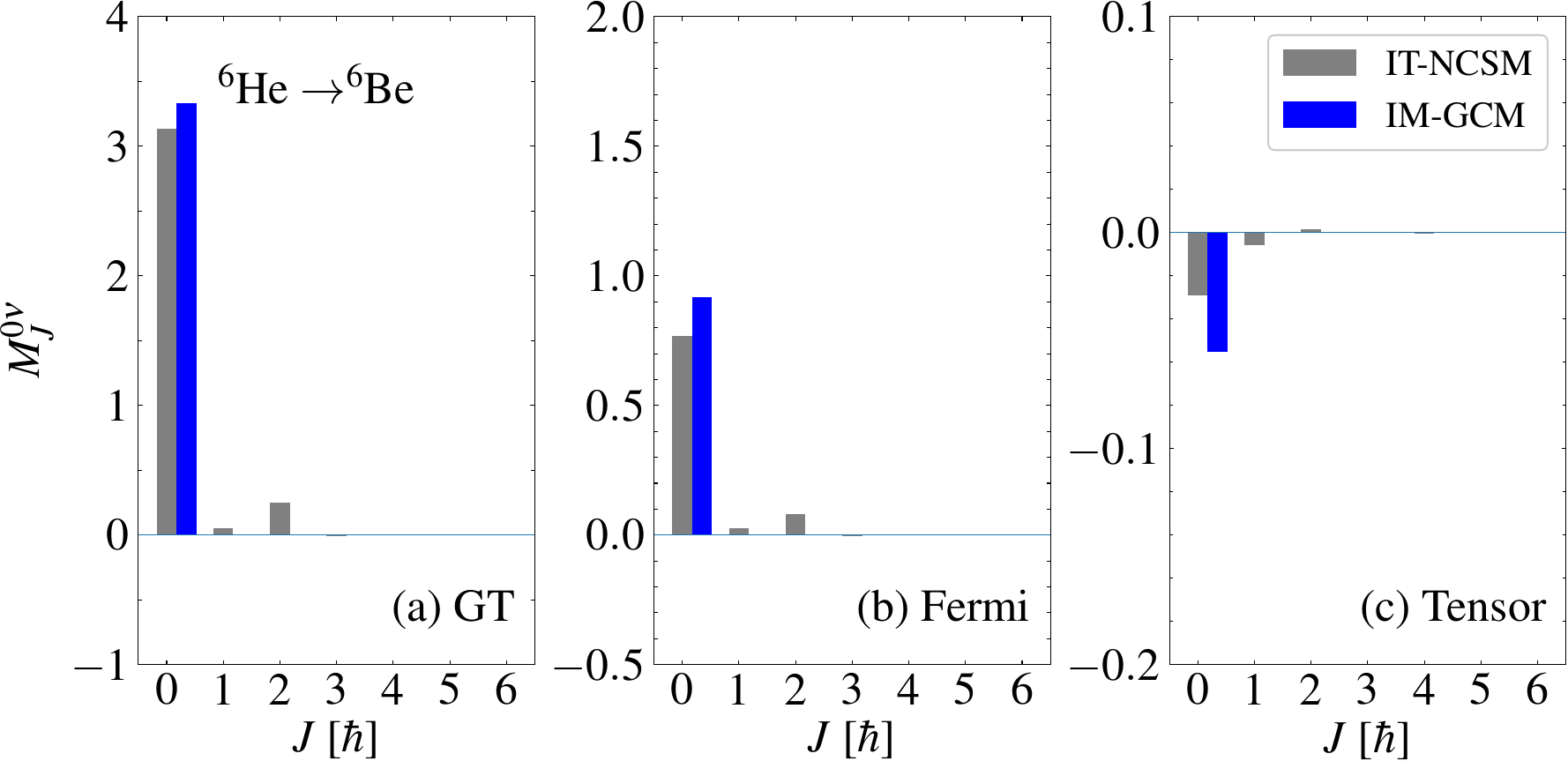}
\caption{\label{fig:He2Be6-MJ}(Color online) The $J$-component $M^{0\nu}_J$ of the  GT, Fermi and tensor parts of the NME for the decay  $^{6}$He $\to$ $^{6}$Be from both IM-GCM and IT-NCSM calculations. }
\end{figure}

 \begin{figure}
\centering 
\includegraphics[width=8.6cm]{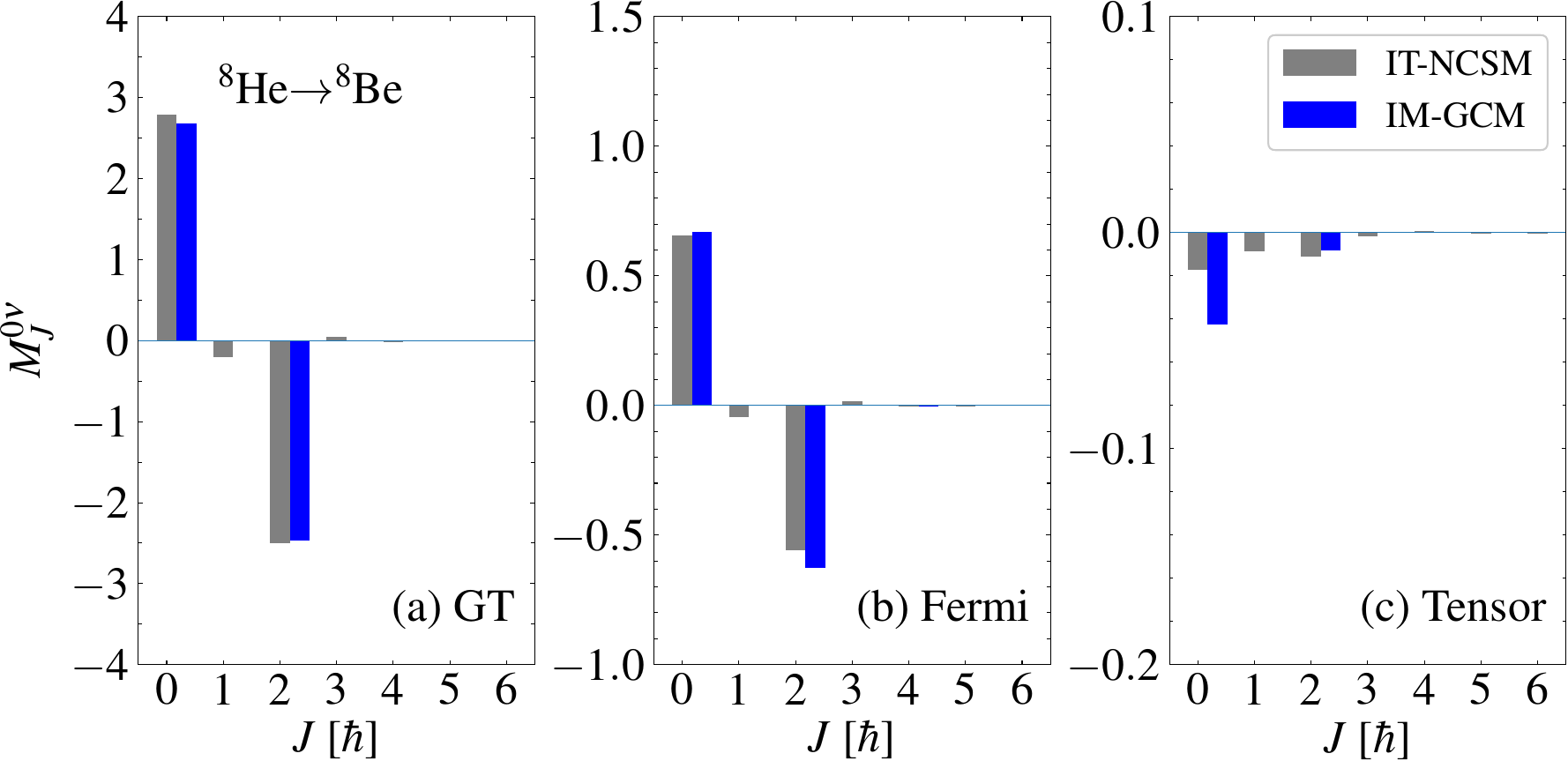}
\caption{\label{fig:He2Be8-MJ}(Color online) Same as in Fig~ \ref{fig:He2Be6-MJ}, but for  the decay $^{8}$He $\to$ $^{8}$Be. }
\end{figure}

 \begin{figure}
\centering 
\includegraphics[width=8.6cm]{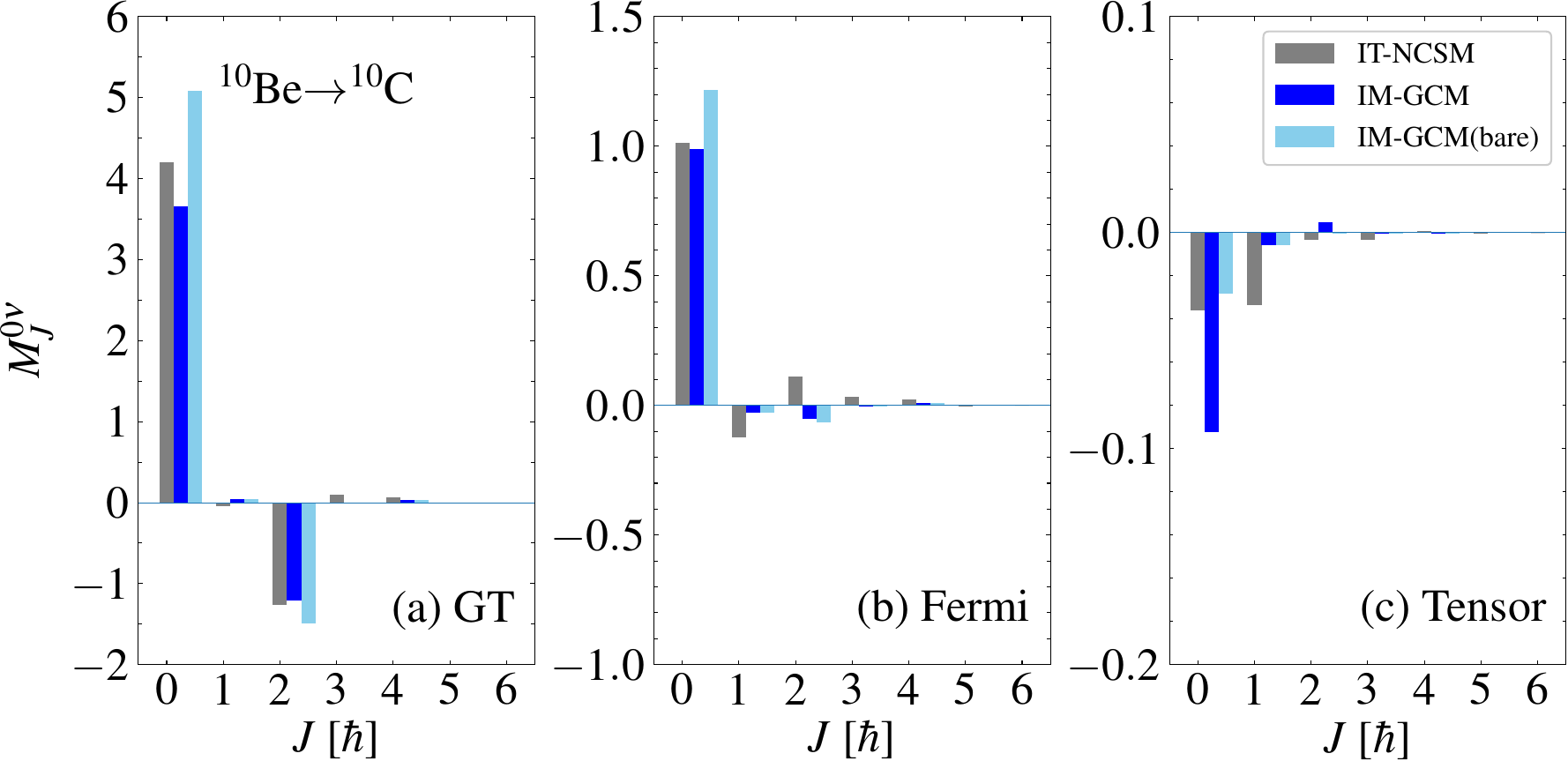}
\caption{\label{fig:Be2C10-MJ}(Color online) The $J$-component $M^{0\nu}_J$ of the  GT, Fermi and tensor parts of the NME for the decay $^{10}$Be $\to$ $^{10}$C from both IM-GCM and IT-NCSM calculations. The results of IM-GCM calculation using the bare transition operator are also given for comparison.
}
\end{figure}

 \begin{figure}
\centering 
\includegraphics[width=8.6cm]{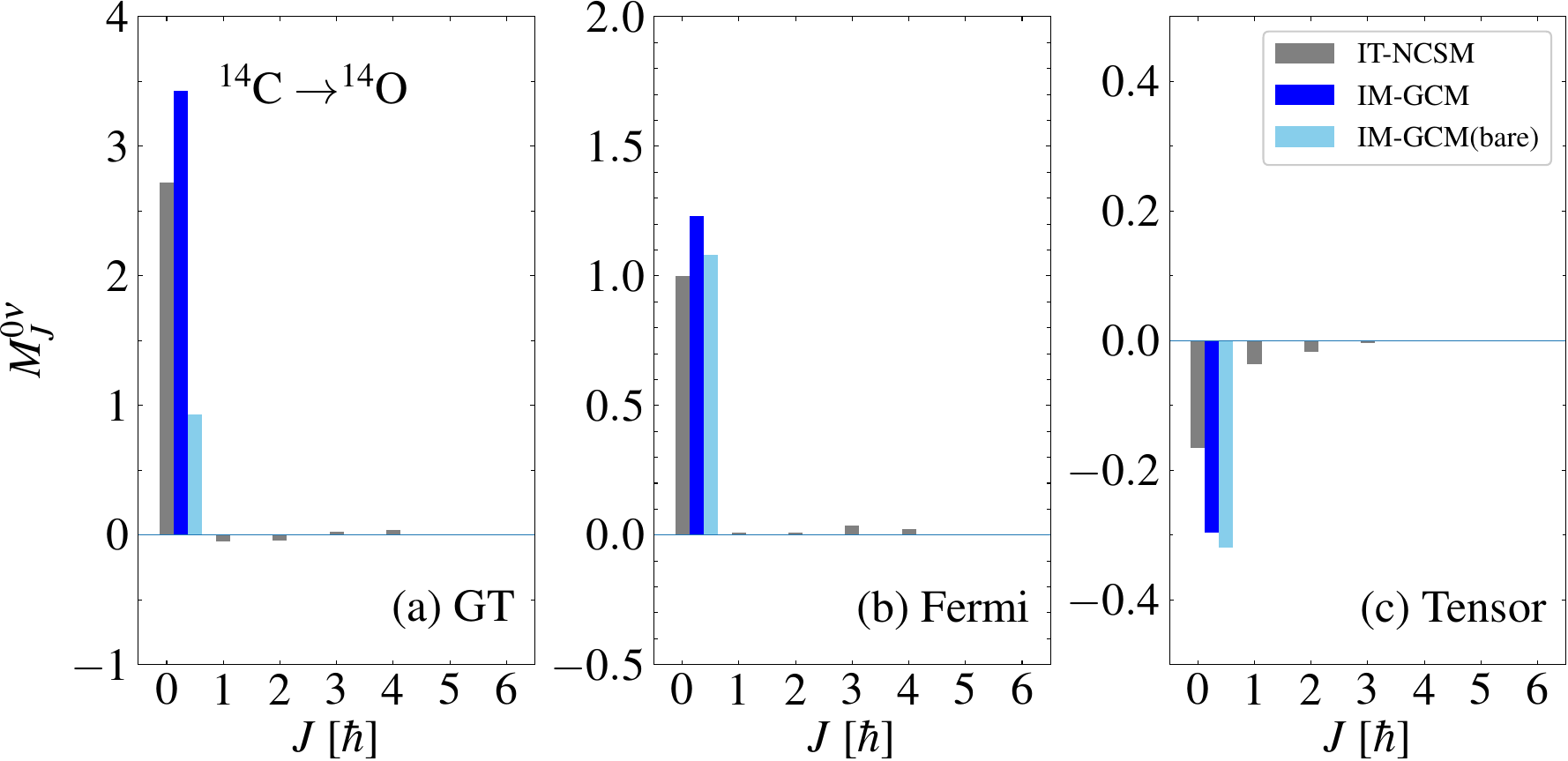}
\caption{\label{fig:C4O14-MJ}(Color online) Same as in Fig.~\ref{fig:Be2C10-MJ}, but for the decay $^{14}$C $\to$ $^{14}$O.}
\end{figure}

Figure \ref{fig:C2O14TBME} displays the renormalization correction $\tensor*[^J]{\delta\mathbb{O}}{^{pp'}_{nn'}}$  to each of the
normalized  
two-body transition matrix elements  $\tensor*[^J]{\mathbb{O}}{^{pp'}_{nn'}}$ with coupled spin $J=0$ in the IMSRG(2) evolution based on the ensemble reference state of $^{14}$C and $^{14}$O, where
\begin{align}
\tensor*[^J]{\delta\mathbb{O}}{^{pp'}_{nn'}} &= \tensor*[^J]{\overline{\mathbb{O}}}{^{pp'}_{nn'}} - \tensor*[^J]{\mathbb{O}}{^{pp'}_{nn'}}.
\end{align}
 This correction is generally small with a few exceptions. For $^{14}$C, the largest change is shown in $\tensor*[^{J=0}]{(\mathbb{O}^{0\nu}_{\text{GT}})}{^{\pi p_{1/2},\pi p_{1/2}}_{\nu p_{1/2},\nu p_{1/2}}}$,
 which is enhanced significantly from $-0.97$ to $-3.57$.
 The enhancement is driven by the contributions of the $pp$ and $hh$ terms in Eq.~\eqref{eq:trans_op_2B} related to the elements $\tensor*[^{J=0}]{\Omega}{^{\pi p_{1/2},\pi p_{1/2}}_{\pi p_{3/2},\pi p_{3/2}}}$
 and $\tensor*[^{J=0}]{\Omega}{^{\nu p_{1/2},\nu p_{1/2}}_{\nu p_{3/2},\nu p_{3/2}}}$ of the Magnus operator.
 
 A similar phenomenon is observed for the $^{10}$Be case:
 The largest change occurs in the two-body transition matrix element
 $\tensor*[^{J=0}]{(\mathbb{O}^{0\nu}_{\text{GT}})}{^{\pi p_{3/2},\pi p_{3/2}}_{\nu p_{3/2},\nu p_{3/2}}}$
 which is reduced from $-3.12$ to $-2.02$.
 In contrast to $^{14}$C, this change is mainly contributed from the $ph$ term in \eqref{eq:trans_op_2B} related to the Magnus operator matrix element
 $\tensor*[^{J=0}]{\Omega}{^{\pi p_{1/2},\nu p_{1/2}}_{\pi p_{3/2},\nu p_{3/2}}}$.
 
 The renormalization effects can be understood as follows.
 The IMSRG flow in the IM-GCM calculation decouples the ground state from excitations and generates large nonzero values for the two-body matrix elements of the $\Omega$ operator connecting the single-particle states below and above the Fermi surface, which are  $p_{3/2}$ and $p_{1/2}$ states respectively in the case of using the ensemble reference states of $^{10}$Be-$^{10}$C and  $^{14}$C-$^{14}$O.
 This large  renormalization effect on a particular two-body matrix element must be compensated by the changes in nuclear wavefunctions.
 However, the IM-GCM presently underestimates (overestimates) the NME for $^{10}$Be ($^{14}$C) because of the adopted truncations in the  IMSRG(2) and GCM parts of our calculations as discussed before. Again as discussed in the \hyperref[appendix]{appendix}, there is about \SI{10}{\percent} reduction contributed from the induced normal-ordered three-body transition operator and \SI{5}{\percent} reduction from the inclusion of neutron-proton isoscalar pairing fluctuations in the GCM calculation.

Finally, we compare the $J$-component $M^{0\nu}_J$ of the  GT, Fermi and tensor parts of the NMEs for  $^{6}$He, $^{8}$He,  $^{10}$Be and  $^{14}$C from both the IT-NCSM and IM-GCM calculations in  Figs.~\ref{fig:He2Be6-MJ}, \ref{fig:He2Be8-MJ}, \ref{fig:Be2C10-MJ}, and \ref{fig:C4O14-MJ}, respectively. 
We find a remarkable agreement in the distribution of the $M^{0\nu}_J$, even though the NMEs from the IT-NCSM calculation tend to be slightly more fragmented than those from the IM-GCM calculation. The NMEs for the $\Delta T=0$ transitions in $^{6}$He and $^{14}$C are determined almost purely by the $J=0$ component.  For others, the NMEs are dominated by the cancellation of $J=0$ and $J=2$ components. As discussed in Ref.~\cite{Yao:2020} with the IM-GCM, the $J=2$ component usually reflects quadrupole deformation effect. The predicted large $J=2$ component for the transition from $^{10}$Be to $^{10}$C is consistent with the observed strong electric quadrupole transition in $^{10}$Be \cite{McCutchan:2009}. Figures \ref{fig:He2Be8-MJ}, and ~\ref{fig:Be2C10-MJ} also display the $M^{0\nu}_J$ by the bare transition operator for $^{10}$Be and  $^{14}$C. It is shown again that the renormalization effect brings the $M^{0\nu}_J$ in the IM-GCM closer to the results of IT-NCSM.

  \section{Summary}
 \label{summary}
 Significant progress has been made in the modeling of the NMEs  for $0\nu\beta\beta$ decays from first principles in recent years. This achievement is mainly attributed to the tremendous development of \emph{ab initio} methods for atomic nuclei with systematically improvable approximations and the use of nuclear Hamiltonians that are softened with SRG transformations. The validation of the impact of these approximations on the NMEs for $0\nu\beta\beta$ decays is an essential step towards the quantification of the uncertainties in theoretically predicted NMEs of candidate $0\nu\beta\beta$ decays.
 
 In this paper, we have presented \emph{ab initio} calculations of the ground-state energies and the NMEs of both $\Delta T=0$ and $\Delta T=2$ $0\nu\beta\beta$ decays in a set of light nuclei with mass number ranging from $A=6$ to $A=22$ with the IM-GCM, VS-IMSRG, and IT-NCSM, starting from the same chiral NN+3N interaction and the same weak transition operator derived from standard light-Majorana neutrino-exchange mechanism. The results have been discussed in comparison with the recently reported results with the NCSM and CCSDT1 \cite{Novario2020}. Our findings are summarized as follows: 
 \begin{itemize}
     \item
     The ground-state energies of all the model calculations agree reasonably well with data.
     The discrepancy from data is within \SI{0.3}{\MeV} per nucleon, except for $\nuclide[8]{Be}$, which has a pronounced 2$\alpha$ cluster structure in its ground state that is challenging to describe in the VS-IMSRG(2). 
     
     \item The NMEs of the $0\nu\beta\beta$ decays with $\Delta T=0$ are generally located around 4.0.
     It turns out that the discrepancy of the NMEs among model predictions can be reduced to be less than \SI{10}{\percent} if the induced normal-ordered three-body transition operator (and neutron-proton isoscalar pairing) is considered in the IM-GCM. 
     
     \item An accurate description of the NMEs for the $\Delta T=2$ transitions is challenging as the values are about one order of magnitude smaller than those for  the $\Delta T=0$ transitions. The NMEs for the transitions from $\nuclide[8]{He}$ to $\nuclide[8]{Be}$ and from $\nuclide[22]{O}$ to $\nuclide[22]{Ne}$ exhibit a similar feature as what been found for the NME of candidate transition from $\nuclide[48]{Ca}$ to $\nuclide[48]{Ti}$ \cite{Yao:2020,Belley2020,Novario2020}, i.e., the NMEs of IM-GCM, VS-IMSRG and IT-NCSM are sandwiched by the upper and lower boundary values from the CCSDT1 calculations with the choice of initial and final (deformed) state as reference state, respectively.  A fairly good agreement is shown in the results of IM-GCM, VS-IMSRG and IT-NCSM for $\nuclide[22]{O}$, and in the results of IM-GCM, IT-NCSM, NCSM, and (the lower boundary value of) CCSDT1 for $\nuclide[8]{He}$.

 \end{itemize}
     
   In short, the present benchmark study provides evidence that the discrepancy between NMEs for $0\nu\beta\beta$ computed with different \emph{ab initio} methods, but using the same input, can provide a meaningful estimate of the truncation errors of the many-body methods.

\begin{acknowledgments}

We thank P.~Gysbers, G.~Hagen, P.~Navratil, S. J. Novario and S.~Quaglioni  for providing us their results of NCSM and CCSDT1 calculations, B.~Bally,  J.~Engel, A. Marquez Romero and T.~R.~Rodríguez for fruitful discussions, and K.~Hebeler for providing us with momentum-space inputs and benchmarks during the construction of our three-nucleon matrix elements. This work is supported in part by the U.S.~Department of Energy, Office of Science, Office of Nuclear Physics under Grants No.\ de-sc0017887 and de-sc0018083 (NUCLEI SciDAC Collaboration), DE-FG02-97ER41019, and DE-SC0015376 (the DBD Topical Theory Collaboration), NSERC, the Arthur B.\ McDonald Canadian Astroparticle Physics Research Institute, the Canadian Institute for Nuclear Physics, and the U.S.~Department of Energy (DOE) under contract DE-FG02-97ER41014. TRIUMF receives funding via a contribution through the National Research Council of Canada. The IM-GCM and IT-NCSM calculations were carried out using the computing resources provided by the Institute for Cyber-Enabled Research at Michigan State University, and the U.S.~National Energy Research Scientific Computing Center (NERSC), a DOE Office of Science User Facility supported by the Office of Science of the U.S.~Department of Energy under Contract No.\ DE-AC02-05CH11231.   The VS-IMSRG computations were performed with an allocation of computing resources on Cedar at WestGrid and Compute Canada, and on the Oak Cluster at TRIUMF managed by the University of British Columbia department of Advanced Research Computing (ARC).
\end{acknowledgments}
 
\appendix*
 
\section{The induced three-body \texorpdfstring{$0\nu\beta\beta$}{0νββ} transition operator}
\label{appendix}

\begin{figure}
\centering
\includegraphics[width=7cm]{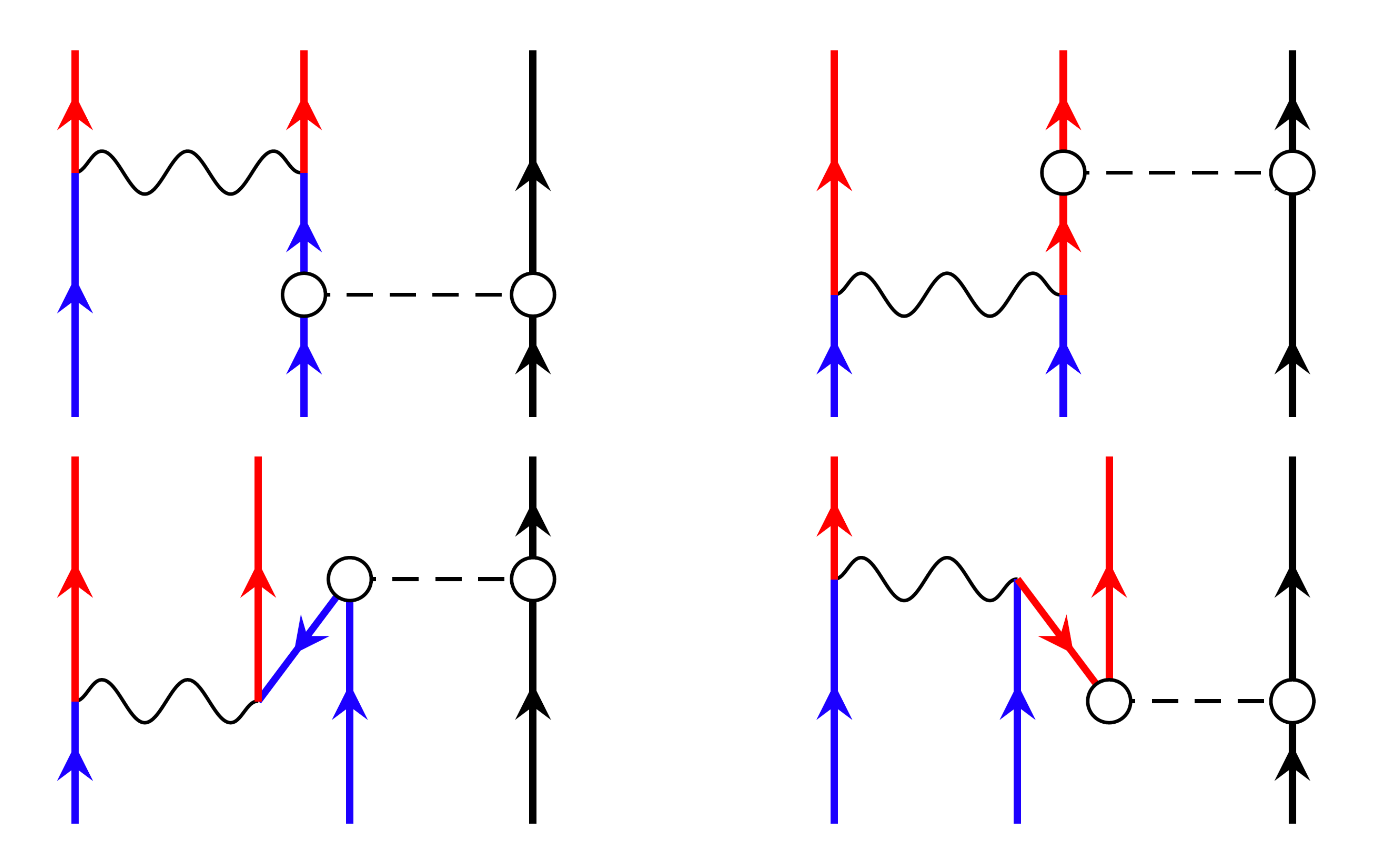} 
\caption{(Color online) Antisymmetrized Goldstone diagrams for the induced three-body transition operator from the commutator $T^{(3)}=[\Omega^{(2)}, O^{0\nu}]^{(3)}$, where dashed lines with hollow dots are for the $\Omega^{(2)}$ in the unitary transformation operator $e^{\Omega}$, wavy lines correspond to $O^{0\nu}$ for the bare $0\nu\beta\beta$ transition operator. The blue, red and black lines with arrows are for neutrons (n), protons (p) and nucleons ($\tau=\mathrm{n/p}$), respectively.
}
\label{fig:induced3B}
\end{figure}

\begin{figure}
\centering
\includegraphics[width=8.6cm]{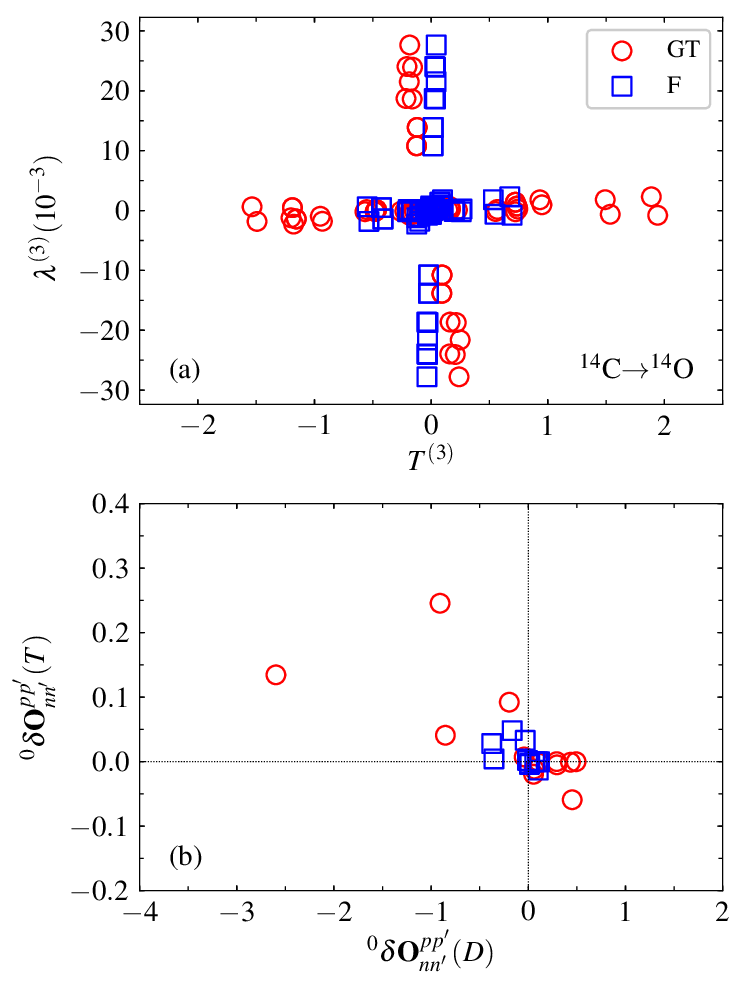} 
\caption{(Color online) (a) The $J$-coupled matrix elements of irreducible three-body transition density $\lambda^{(3)}$ against those of induced three-body GT, Fermi and tensor transition operator  $T^{(3)}_\alpha=[\Omega^{(2)}, O^{0\nu}_\alpha]^{(3)}$ in the IM-GCM for $\nuclide[14]{C}$, where $\alpha$ labels GT and Fermi respectively. (b) Correction to the normalized matrix elements of two-body transition operator $O^{0\nu}$ from $T^{(3)}$ against that from $D^{(2)}$ under NO2B approximation. See text for details.}
\label{fig:ME3B}
\end{figure}

In this section, we assess the corrections to the NME in IM-GCM calculation from the induced three-body transition operator in IMSRG(2), which is composed of NO2B term and normal-ordered three-body (NO3B) term. The NO2B part of  the induced three-body transition operator has already been taken into account in the NME. Here we will evaluate the contributions from the NO3B part of the induced transition operator. The consideration of these contributions in principle requires the extension of the IMSRG(2) to IMSRG(3), in which all the operators are truncated up to NO3B terms. This extension is however still a formidable computational challenge \cite{Hergert:2018th}. Therefore, we only examine what are expected to be the predominant contributions from the NO3B transition operator. 
To this end, we rewrite the IMSRG-evolved transition operator  in terms of the rank of each operator
\begin{align}
\label{eq:append1}
   e^{\Omega}O^{0\nu}e^{-\Omega} 
   &= O^{0\nu} + [\Omega, O^{0\nu}]^{(2)} + \tfrac1{2!} [\Omega, [\Omega, O^{0\nu}]^{(2)}]^{(2)} + \dotsb \nonumber\\
    &\hphantom{{}={}} + [\Omega, O^{0\nu}]^{(3)} + \tfrac1{2!}[\Omega, [\Omega, O^{0\nu}]^{(3)}]^{(2)} + \dotsb
\end{align}
where the first line collects all the two-body operators that are included in the IMSRG(2) framework.

The three-body terms in $[\Omega, O^{0\nu}]$ are generated by $\Omega^{(2)}$, so we introduce the notation $ T^{(3)} $ for the induced NO3B transition operator  as shown diagrammatically in Fig.~\ref{fig:induced3B},
   \beqn
   \label{eq:3B}
    T^{(3)} 
    &=&[\Omega^{(2)}, O^{0\nu}]^{(3)}
    =\dfrac{1}{36}\sum T^{pp'\tau}_{nn'\tau'} \nord{A^{pp'\tau}_{nn'\tau'}}.
    \eeqn
The antisymmetrized three-body matrix element $T^{abc}_{def}$ in natu\-ral-orbital basis is given by
\beq
 T^{abc}_{def}  
 = \sum_g\mathcal{A}(\Omega^{ag}_{ef} O^{bc}_{dg}  - O^{ag}_{ef}  \Omega^{bc} _{dg})\,,
\eeq
which is composed of 9 terms given the fact that the antisymmetry under exchange of $b$ and $c$ (and of $e$ and $f$) is already built into $\Omega$ and $O$.
The correction to the NME due to $T^{(3)}$ reads
   \begin{align}
   \label{direct}
   \delta M^{0\nu}(T^{(3)}) 
   &= \dfrac{1}{36}\sum T^{pp'\tau}_{nn'\tau'}  \braket{ \Psi_F | \nord{A^{pp'\tau}_{nn'\tau'}} | \Psi_I } \nonumber\\
   &= \dfrac{1}{36}\sum T^{pp'\tau}_{nn'\tau'}  \lambda^{pp'\tau}_{nn'\tau'},
   \end{align}
which depends on the irreducible three-body transition density $\lambda^{(3)}$, 
   \beqn
   \label{lambda3B}
    \lambda^{pp'\tau}_{nn'\tau'}  =  \rho^{pp'\tau}_{nn'\tau'} -  \mathcal{A}(\rho^{pp'}_{nn'} \lambda^\tau_{\tau'}).
   \eeqn
In the above expression, $\rho^{pp'\tau}_{nn'\tau'}$ and $\rho^{pp}_{nn'}$ are the three- and two-body transition density, and $\lambda^\tau_{\tau'}$ is the one-body density of the reference state, which conserves charge and isospin. The formulas for the (irreducible) three-body (transition) density can be found in Refs.~\cite{Yao:2018wq,Wang2018}.
Since the irreducible three-body transition density is generally small, the correction $\delta M^{0\nu}(T^{(3)})$ is expected to be small as well.

For illustration, we compute $T^{(3)}$ and $\delta M^{0\nu}(T^{(3)})$ for $\nuclide[14]{C}$ within the \emph{spsd} model space, which is composed of six orbitals of neutrons and protons: $0s_{1/2}, 0p_{3/2}, 0p_{1/2}, 1s_{1/2}, 0d_{5/2}$, and $0d_{3/2}$, where the predominant matrix elements of one-body density, two-body transition density, and the irreducible three-body transition density live.

Figure~\ref{fig:ME3B}(a) displays the $J$-coupled matrix elements of $\tensor*{\lambda}{^{(3)}}$ with magnitude greater than $10^{-4}$ against those of $T^{(3)}$.
The $\tensor*{\lambda}{^{(3)}}$ matrix elements are on the order of $10^{-2}$ while those of the induced three-body GT transition operator range from $-1.5$ to $2.0$. Compared to the IMSRG(2) renormalization effect on the two-body part of the transition operator (cf.\ Fig.~\ref{fig:C2O14TBME}), which changes the unnormalized two-body matrix elements by up to 5.2, this is a \SI{38}{\percent} effect.
However, in addition to their small magnitude, the distributions of the three-body matrix elements $\lambda^{(3)}$ are misaligned with those of $T^{(3)}$, so that large correlation matrix elements get multiplied by small transition matrix elements and \emph{vice versa} when computing the NME.
This leads to a very small correction $\delta M^{0\nu}(T^{(3)}) \approx \num{-2e-3}$ for the GT transition. The contribution to the Fermi and tensor parts is even smaller.

Next, we examine the contribution of the second term in the second line of \eqref{eq:append1}.
This term modifies the two-body transition operator and it is given by
\begin{align}
   \label{eq:3B2B2B}
   \delta O^{0\nu}
   &=\dfrac{1}{2!}[\Omega^{(2)},[\Omega^{(2)},O^{0\nu}]^{(3)}]^{(2)}\nonumber \\
   &=\dfrac{1}{2!}[\Omega^{(2)},T^{(3)}]^{(2)}\nonumber \\
   &\equiv \dfrac{1}{4} \sum_{abcd} \delta O^{ab}_{cd} \nord{A_{{cd}}^{{ab}}}
\end{align}
Considering that the irreducible two-body density $\lambda^{(2)}$ is generally much smaller than the one-body density, it is expected to be a good approximation to consider the terms that depend only on the one-body density.
Under this approximation, we find
 \begin{align}
    \delta O^{ab}_{cd}  
    &\simeq \dfrac{1}{4}\sum_{efg}  \Bigg(
    \Omega^{be}_{fg}T^{fga}_{cde}  
    +\Omega^{de}_{fg}T^{fgc}_{abe}
    - \Omega^{ae}_{fg}T^{fgb}_{cde}  
    - \Omega^{ce}_{fg}T^{fgd}_{abe}  
    \Bigg) \nonumber\\
    &
     \times(n_e \bar{n}_f \bar{n}_g + \bar{n}_e n_f n_g),  
    \end{align} 
where $\bar n_i = 1 - n_i$ with $n_i \in [0,1]$ being the occupation number of the $i$-th single-particle state. Using the \texttt{AMC} package \cite{Tichai:2020}, we find the $J$-coupled expression for the unnormalized two-body matrix element 
\begin{widetext}
 
 \beqn
        \tensor*[^{J_0}]{\delta O}{^{ab}_{cd}} 
         &=& \dfrac{1}{4}\sum_{efg} 
         (n_e \bar{n}_f \bar{n}_g + \bar{n}_e n_f n_g)
         \sum_{J_1J_{123}}\hat J^{-1}_0\hat J_1 \hat J^2_{123}
          (-1)^{J_1+j_e+1} \nonumber\\
            &&\times
         \Bigg( 
          (-1)^{j_b} 
            \Gj{j_a}{j_b}{J_0}{j_e}{J_{123}}{J_1}  
            \tensor*[^{J_1}]{\Omega}{^{be}_{fg}} 
             \tensor*[^{J_{123}}]{T}{^{(fg)J_1,a}_{(cd)J_0,e}} 
            +  (-1)^{j_d}
            \Gj{j_c}{j_d}{J_0}{j_e}{J_{123}}{J_2}  
            \tensor*[^{J_1}]{\Omega}{^{de}_{fg}} 
             \tensor*[^{J_{123}}]{T}{^{(fg)J_1,c}_{(ab)J_0,e}}\nonumber\\
            &&
              +  (-1)^{J_0+j_b}
            \Gj{j_b}{j_a}{J_0}{j_e}{J_{123}}{J_1}  
            \tensor*[^{J_1}]{\Omega}{^{ae}_{fg}} 
             \tensor*[^{J_{123}}]{T}{^{(fg)J_1,b}_{(cd)J_0,e}}
              +  (-1)^{J_0+j_d}
            \Gj{j_d}{j_c}{J_0}{j_e}{J_{123}}{J_1}  
            \tensor*[^{J_1}]{\Omega}{^{ce}_{fg}} 
             \tensor*[^{J_{123}}]{T}{^{(fg)J_1,d}_{(ab)J_0,e}} 
             \Bigg).  
 \eeqn
\end{widetext}
where the $\tensor*[^{J_{123}}]{T}{^{(ab)J_0,g}_{(de)J_1,f}}$ is a $J$-coupled form of matrix element $T^{abg}_{def}$. We evaluate $\tensor*[^{J_0}]{\delta O}{^{ab}_{cd}}$ for the transition from $\nuclide[14]{C}$ to $\nuclide[14]{O}$ and find two largest GT matrix elements which are $\tensor*[^{J=0}]{\delta O}{^{\pi p_{1/2},\pi p_{1/2}}_{\nu p_{1/2},\nu p_{1/2}}}=0.27$ and $\tensor*[^{J=0}]{\delta O}{^{\pi p_{3/2},\pi p_{3/2}}_{\nu p_{1/2},\nu p_{1/2}}}=0.49$, respectively, as shown in Fig.~\ref{fig:ME3B}(b).
The size of these matrix elements is an order of magnitude smaller than those of $T^{(3)}$, as is expected for a higher-order correction.
Since it modifies the two-body transition operator, the contributions from $\delta O^{0\nu}$ are not suppressed by the smallness of $\lambda^{(3)}$ and can yield a sizable correction to the NME,
\begin{align}
   \label{eq:3B2B2BNME}
   \delta M^{0\nu}(\delta O) 
   &= \dfrac{1}{4}\sum_{pp'nn'} \delta O^{pp'}_{nn'}  \rho^{pp'}_{nn'}.
\end{align}
This correction reduces the NME by 0.263 and 0.094 for the GT and Fermi, respectively. The total NME for $\nuclide[14]{C}$ by the IM-GCM becomes 4.01,  in better agreement with the value 4.03 of VS-IMSRG (which of course has its own $T^{(3)}$ corrections to be considered), and \SI{11}{\percent} larger than the IT-NCSM value of 3.55.  The latter difference can be reduced further, to about \SI{6}{\percent}, by including neutron-proton isoscalar pairing fluctuations in the GCM calculation.

 \bibliographystyle{apsrev4-1} 
 
%


\end{document}